\documentclass[preprint,prd,aps,showpacs,showkeys,nofootinbib]{revtex4}
\usepackage{graphicx}
\begin{document}


\title{GKZ-hypergeometric systems for Feynman integrals}

\author{Tai-Fu Feng\footnote{email:fengtf@hbu.edu.cn}$^{a,b,g}$,
Chao-Hsi Chang\footnote{email:zhangzx@itp.ac.cn}$^{c,d,e}$,
Jian-Bin Chen$^{f}$\footnote{email:chenjianbin@tyut.edu.cn},
Hai-Bin Zhang$^{a,b}$\footnote{email:hbzhang@hbu.edu.cn}}

\affiliation{$^a$Department of Physics, Hebei University, Baoding, 071002, China}
\affiliation{$^b$Hebei Key Laboratory of High-precision Computation and Application of Quantum Field Theory, Baoding, 071002, China}
\affiliation{$^c$Key Laboratory of Theoretical Physics, Institute of Theoretical Physics,
Chinese Academy of Science, Beijing, 100190, China}
\affiliation{$^d$CCAST (World Laboratory), P.O.Box 8730, Beijing, 100190, China}
\affiliation{$^e$School of Physical Sciences, University of Chinese
Academy of Sciences, Beijing 100049, China}
\affiliation{$^f$Department of Physics, Taiyuan University of Technology, Taiyuan, 030024, China}
\affiliation{$^g$Department of Physics, Chongqing University, Chongqing, 401331, China}

\begin{abstract}
Basing on the systems of linear partial differential equations derived from Mellin-Barnes
representations and Miller's transformation, we obtain GKZ-hypergeometric systems
of one-loop self energy, one-loop triangle, two-loop vacuum, and
two-loop sunset diagrams, respectively. The codimension of derived GKZ-hypergeometric system
equals the number of independent dimensionless ratios among the external momentum
squared and virtual mass squared.
Taking GKZ-hypergeometric systems of one-loop self energy, massless one-loop triangle,
and two-loop vacuum diagrams as examples, we present in detail how to perform triangulation
and how to construct canonical series solutions in the corresponding convergent regions.
The series solutions constructed for these hypergeometric systems recover
the well known results in literature.
\end{abstract}


\maketitle

\section{Introduction\label{sec1}}
\indent\indent
A central target for particle physics now is to test the standard model (SM) and to search for
new physics (NP) beyond the SM \cite{CEPC-SPPC,ILC,HI-LHC} after the
discovery of the Higgs boson~\cite{CMS2012,ATLAS2012}. In order to predict
the electroweak observables precisely with dimensional regularization~\cite{tHooft1979,Passarino1979},
one should evaluate the Feynman integrals exactly in the time-space dimension
$D=4-2\varepsilon$ at first. Nevertheless each method presented in Ref.~\cite{V.A.Smirnov2012}
has its blemishes since it can only be applied to the Feynman diagrams with
special topologies and kinematic invariants.

It was proposed long ago to consider Feynman integrals as the generalized hypergeometric
functions~\cite{Regge1967}. Certainly Feynman integrals satisfy indeed the systems of
holonomic linear partial differential equations (PDEs)~\cite{Kashiwara1976} whose singularities
are determined by the Landau singularities. Recently the author of Ref.~\cite{Nasrollahpoursamami2016}
shows that the $D-$module of a Feynman diagram is isomorphic to
Gel'fand-Kapranov-Zelevinsky (GKZ) $D-$module~\cite{Gelfand1987,Gelfand1988,Gelfand1988a,Gelfand1989,Gelfand1990}.

Some Feynman integrals are already expressed as the hypergeometric series in the corresponding
parameter space.
In Ref.~\cite{Davydychev1} the massless $C_{_0}$ function
is presented as the linear combination of the fourth kind of Appell function
$F_{_4}$ whose arguments are the dimensionless ratios among the external momentum squared,
and is simplified further as the linear combination of the Gauss function $_2F_1$
through the quadratic transformation~\cite{Davydychev1993NPB} in the literature~\cite{Davydychev2000}.
With some special assumptions on the virtual masses,
the analytic expressions of the scalar integral $C_{_0}$ are given by
the multiple hypergeometric functions in Ref.~\cite{Davydychev3} through
Mellin-Barnes representations. Taking the massless $C_{_0}$ function as
an example, the author of Ref~\cite{Davydychev1992JPA} presents an algorithm
to evaluate the scalar integrals of one-loop vertex-type Feynman diagrams.
Certainly, some analytic results of the $C_{_0}$ function can also be extracted from
the expressions for the scalar integrals of one-loop massive $N-$point
Feynman diagrams~\cite{Davydychev1991JMP,Davydychev1992JMP}. Feynman parametrization
and Mellin-Barnes contour integrals can be applied to evaluate Feynman integrals
of ladder diagrams with three or four external lines~\cite{Davydychev1993}. In addition,
the literature~\cite{Davydychev2006} also provides a geometrical interpretation
of the analytic expressions of the scalar integrals from one-loop $N-$point
Feynman diagrams. Using the recurrence relations respecting the time-space
dimension, the authors of Refs.~\cite{Tarasov2000,Tarasov2003} formulate one-loop
two-point function $B_{_0}$ as the linear combination of the Gauss function
$_2F_1$, one-loop three-point function $C_{_0}$
with arbitrary external momenta and virtual masses as the linear combination
of the Appell function $F_{_1}$,
and one-loop four-point function $D_{_0}$ with arbitrary external momenta
and virtual masses as the linear combination of the Lauricella-Saran function $F_{_s}$
with three arguments, respectively. The expression for the scalar integral $C_{_0}$ is convenient
for analytic continuation and numerical evaluation because continuation of the Appell
functions has been analyzed thoroughly. Nevertheless, how to perform continuation
of the Lauricella-Saran function $F_{_s}$ outside its convergent domain is still a challenge.
Expressing the relevant Feynman integral as a linear combination of generalized hypergeometric
functions in dimension regularization,
the authors of Ref.~\cite{Kalmykov2009} analyze Laurent expansion of these hypergeometric
functions around $D=4$. The differential-reduction algorithm to evaluate those hypergeometric
functions can be found in Refs.~\cite{Bytev2010,Kalmykov2011,Bytev2013,Bytev2015,Bytev2016}.
A hypergeometric system of linear PDEs is given through Mellin-Barnes
representation~\cite{Kalmykov2012}, where the system of linear PDEs is satisfied
by the corresponding Feynman integral in the whole parameter space.
Some irreducible master integrals for sunset and bubble Feynman diagrams with
generic values of masses and external momenta are explicitly evaluated via their
Mellin-Barnes representations in Ref.~\cite{Kalmykov2017}.

Taking some special assumptions on the virtual masses and external momenta, the author of
Ref.~\cite{Cruz2019} presents some GKZ-hypergeometric systems of Feynman integrals with codimension$=0$
or codimension$=1$ through Lee-Pomeransky parametric representations~\cite{Lee2013}.
Using the triangulations of the Newton polytope of Lee-Pomeransky polynomial,
the author of Ref.~\cite{Klausen2019} presents GKZ-hypergeometric system of
the sunset diagram of codimension$=6$. He also constructs canonical series solutions
which contain three redundant variables besides three independent dimensionless ratios among
the external momentum squared $p^2$ and three virtual mass squared $m_{_i}^2\;\;(i=1,\;2,\;3)$
under the assumption $|p^2-m_{_1}^2-m_{_2}^2-m_{_3}^2|\gg m_{_i}^2$. Actually it
is a common defect of GKZ-hypergeometric systems originating from Lee-Pomeransky polynomial
of the corresponding Feynman diagrams that codimension is far larger than the number of
independent dimensionless ratios among the external momentum squared and virtual mass squared.
To construct canonical series solutions with suitable independent variables,
one should compute the restricted $D$-module of GKZ-hypergeometric system originating from
Lee-Pomeransky representations on corresponding hyperplane in the parameter
space~\cite{Oaku1997,Walther1999,Oaku2001}.

Some holonomic systems of linear PDEs are given through
Mellin-Barnes representations of the concerned Feynman integrals in
Refs.\cite{Kalmykov2012,Feng2018,Feng2019}.
Following the work of W.~Miller~\cite{Miller68,Miller72},
one derives GKZ-hypergeometric system of Feynman integrals,
whose codimension equals the number of independent dimensionless ratios among
the external momentum squared and virtual mass squared. Using those holonomic
systems given in Refs.\cite{Kalmykov2012,Feng2018,Feng2019}, we present here relevant
GKZ-hypergeometric systems for Feynman integrals of one-loop self-energy,
two-loop vacuum, two-loop sunset, and one-loop triangle diagrams.
Taking Feynman integrals of one-loop self-energy,
two-loop vacuum, and massless one-loop triangle diagrams as examples, we illuminate
how to construct canonical series solutions from those relevant GKZ-hypergeometric
systems~\cite{M.Saito2000}, and how to derive the convergent regions of those series
with Horn's study~\cite{Horn1889}. To shorten the length of text,
we don't state those mathematical concepts and theorems that have been used in our analyses here,
because they can be found in some well-known mathematical textbooks~\cite{M.Saito2000,M.E.Taylor12,
L.J.Slater66,J.Leray1959,Cox1991,Cox1998,Sturmfels1995,Eisenbud1995,Coutinho1995}.
Basing on Mellin-Barnes representations of one-loop Feynman diagrams
or those multiloop diagrams with two vertices, we can derive GKZ-hypergeometric
systems through Miller's transformation, whose codimension of GKZ-hypergeometric system
equals the number of independent dimensionless ratios among the external momentum squared
and virtual mass squared. Using toric geometry and mirror symmetry, one also derives
the PDEs satisfied by Feynman integrals of the multiloop sunset diagrams~\cite{Vanhove2018}.
Nevertheless for generic multiloop Feynman diagrams
such as that presented in Refs~\cite{V.A.Smirnov1999,J.B.Tausk1999}, the corresponding codimension
of GKZ-hypergeometric system derived is far larger than the number of independent
dimensionless ratios, whether using Mellin-Barnes or Lee-Pomeransky representations. In order to construct
canonical series solutions properly, the corresponding GKZ-hypergeometric
system is restricted to the hyperplane in parameter space.

The generally strategy for analyzing Feynman integral includes three steps here.
First we obtain the holonomic system of linear PDEs satisfied by corresponding Feynman integral through
its Mellin-Barnes representation, next find GKZ-hypergeometric system via Miller's transformation,
and finally construct canonical series solutions. The integration constants, i.e.
the combination coefficients, are determined by the corresponding Feynman integral with some special
kinematic parameters. To make the analytic continuation of those canonical series solutions from their convergent regions
to the whole parameter space, one can perform some linear fractional transformations
among the complex variables.

Our presentation is organized as following. Through Miller's transformation,
we derive GKZ-hypergeometric systems of Feynman integrals of one-loop
self energy, massless one-loop triangle, and two-loop vacuum diagrams by using the
holonomic systems of linear PDEs in
Refs.\cite{Kalmykov2012,Feng2018,Feng2019} in section \ref{sec2}.
Then we present in detail how to perform triangulation and how to construct canonical series
solutions from those GKZ-hypergeometric systems in section \ref{sec3}.
Actually some well-known results are recovered with the approach presented here.
In section \ref{sec4}, we present GKZ-hypergeometric systems
for the sunset diagram with three differential masses, $C_{_0}$ function with one nonzero
virtual mass, and $C_{_0}$ function with three differential virtual masses, respectively.
The conclusions are summarized in section \ref{sec5}.

\section{GKZ-hypergeometric systems of one-loop self energy, massless one-loop triangle,
and two-loop vacuum diagrams\label{sec2}}
\indent\indent
Adopting the notation of Refs.~\cite{Feng2018,Feng2019}, we write the scalar integrals of one-loop
self energy, massless one-loop triangle, and two-loop vacuum diagrams respectively as
\begin{eqnarray}
&&B_{_0}(p^2,\;m_{_1}^2,\;m_{_2}^2)={i\over(4\pi)^2}\Big({4\pi\mu^2\over-p^2}\Big)^{2-D/2}
f_{_B}\left(\left.\begin{array}{cc}a_{_B},&b_{_B}\\
c_{_B},&c_{_B}^\prime\end{array}\right|x_{_B},\;y_{_B}\right)
\;,\nonumber\\
&&C_{_0}(p_{_1}^2,\;p_{_2}^2,\;p_{_3}^2)={i\over(4\pi)^2p_{_3}^2}
\Big({4\pi\mu^2\over-p_{_3}^2}\Big)^{2-D/2}
f_{_C}\left(\left.\begin{array}{cc}a_{_C},&b_{_C}\\
c_{_C},&c_{_C}^\prime\end{array}\right|x_{_C},\;y_{_C}\right)
\;,\nonumber\\
&&V_{_2}(m_{_1}^2,\;m_{_2}^2,\;m_{_3}^2)={m_{_3}^2\over(4\pi)^4}
\Big({4\pi\mu^2\over m_{_3}^2}\Big)^{4-D}
f_{_V}\left(\left.\begin{array}{cc}a_{_V},&b_{_V}\\
c_{_V},&c_{_V}^\prime\end{array}\right|x_{_V},\;y_{_V}\right)\;,
\label{GKZ1}
\end{eqnarray}
where $\mu$ denotes the renormalization energy scale, and $x_{_B}=m_{_1}^2/p^2$,
$y_{_B}=m_{_2}^2/p^2$, $x_{_C}=p_{_1}^2/p_{_3}^2$, $y_{_C}=p_{_2}^2/p_{_3}^2$,
$x_{_V}=m_{_1}^2/m_{_3}^2$, $y_{_V}=m_{_2}^2/m_{_3}^2$, respectively.
Here the dimensionless functions $f_{_i}\;(i=B,\;C,\;V)$
comply with the fourth Appell's system of linear PDEs
\begin{eqnarray}
&&\Big\{\hat{\vartheta}_{_{x_{_i}}}(\hat{\vartheta}_{_{x_{_i}}}+c_{_i}-1)-x_{_i}(\hat{\vartheta}_{_{x_{_i}}}
+\hat{\vartheta}_{_{y_{_i}}}+a_{_i})(\hat{\vartheta}_{_{x_{_i}}}+\hat{\vartheta}_{_{y_{_i}}}+b_{_i})\Big\}
f_{_i}\left(\left.\begin{array}{cc}a_{_i},&b_{_i}\\
c_{_i},&c_{_i}^\prime\end{array}\right|x_{_i},\;y_{_i}\right)=0
\;,\nonumber\\
&&\Big\{\hat{\vartheta}_{_{y_{_i}}}(\hat{\vartheta}_{_{y_{_i}}}+c_{_i}^\prime-1)-y_{_i}(\hat{\vartheta}_{_{x_{_i}}}
+\hat{\vartheta}_{_{y_{_i}}}+a_{_i})(\hat{\vartheta}_{_{x_{_i}}}+\hat{\vartheta}_{_{y_{_i}}}+b_{_i})\Big\}
f_{_i}\left(\left.\begin{array}{cc}a_{_i},&b_{_i}\\
c_{_i},&c_{_i}^\prime\end{array}\right|x_{_i},\;y_{_i}\right)=0\;,
\label{GKZ2}
\end{eqnarray}
with the Euler operator $\hat{\vartheta}_{_x}=x\partial_{_x}$.
Correspondingly those parameters in Eq.~(\ref{GKZ2}) are
\begin{eqnarray}
&&a_{_{B,V}}=2-{D\over2}\;,\;\;b_{_{B,V}}=3-D\;,\nonumber\\
&&c_{_{B,V}}=c^\prime_{_{B,V}}=2-{D\over2}\;,\nonumber\\
&&a_{_C}=1\;,\;\;b_{_C}=3-{D\over2}\;,\nonumber\\
&&c_{_C}=c^\prime_{_C}=3-{D\over2}\;.
\label{GKZ3}
\end{eqnarray}
Generally those holonomic systems presented above originate
from Mellin-Barnes representations of the corresponding Feynman integrals\cite{Kalmykov2012,Feng2018,Feng2019}.
Using the systems of linear PDEs in Eq.~(\ref{GKZ2}),
one derives the following relations between $f_{_i}\;(i=B,\;C,\;V)$ and their contiguous
functions
\begin{eqnarray}
&&(\hat{\vartheta}_{_{x_{_i}}}+\hat{\vartheta}_{_{y_{_i}}}+a_{_i})
f_{_i}\left(\left.\begin{array}{cc}a_{_i},&b_{_i}\\
c_{_i},&c_{_i}^\prime\end{array}\right|x_{_i},\;y_{_i}\right)=
a_{_i}f_{_i}\left(\left.\begin{array}{cc}a_{_i}+1,&b_{_i}\\
c_{_i},&c_{_i}^\prime\end{array}\right|x_{_i},\;y_{_i}\right)
\;,\nonumber\\
&&(\hat{\vartheta}_{_{x_{_i}}}+\hat{\vartheta}_{_{y_{_i}}}+b_{_i})
f_{_i}\left(\left.\begin{array}{cc}a_{_i},&b_{_i}\\
c_{_i},&c_{_i}^\prime\end{array}\right|x_{_i},\;y_{_i}\right)=
b_{_i}f_{_i}\left(\left.\begin{array}{cc}a_{_i},&b_{_i}+1\\
c_{_i},&c_{_i}^\prime\end{array}\right|x_{_i},\;y_{_i}\right)
\;,\nonumber\\
&&(\hat{\vartheta}_{_{x_{_i}}}+c_{_i}-1)
f_{_i}\left(\left.\begin{array}{cc}a_{_i},&b_{_i}\\
c_{_i},&c_{_i}^\prime\end{array}\right|x_{_i},\;y_{_i}\right)=
(c_{_i}-1)f_{_i}\left(\left.\begin{array}{cc}a_{_i},&b_{_i}\\
c_{_i}-1,&c_{_i}^\prime\end{array}\right|x_{_i},\;y_{_i}\right)
\;,\nonumber\\
&&(\hat{\vartheta}_{_{y_{_i}}}+c_{_i}^\prime-1)
f_{_i}\left(\left.\begin{array}{cc}a_{_i},&b_{_i}\\
c_{_i},&c_{_i}^\prime\end{array}\right|x_{_i},\;y_{_i}\right)=
(c_{_i}^\prime-1)f_{_i}\left(\left.\begin{array}{cc}a_{_i},&b_{_i}\\
c_{_i},&c_{_i}^\prime-1\end{array}\right|x_{_i},\;y_{_i}\right)
\;,\nonumber\\
&&\partial_{_{x_{_i}}}f_{_i}\left(\left.\begin{array}{cc}a_{_i},&b_{_i}\\
c_{_i},&c_{_i}^\prime\end{array}\right|x_{_i},\;y_{_i}\right)=
{a_{_i}b_{_i}\over c_{_i}}f_{_i}\left(\left.\begin{array}{cc}a_{_i}+1,&b_{_i}+1\\
c_{_i}+1,&c_{_i}^\prime\end{array}\right|x_{_i},\;y_{_i}\right)
\;,\nonumber\\
&&\partial_{_{y_{_i}}}f_{_i}\left(\left.\begin{array}{cc}a_{_i},&b_{_i}\\
c_{_i},&c_{_i}^\prime\end{array}\right|x_{_i},\;y_{_i}\right)=
{a_{_i}b_{_i}\over c_{_i}^\prime}f_{_i}\left(\left.\begin{array}{cc}a_{_i}+1,&b_{_i}+1\\
c_{_i},&c_{_i}^\prime+1\end{array}\right|x_{_i},\;y_{_i}\right)\;.
\label{GKZ4}
\end{eqnarray}
Following the work of W.~Miller~\cite{Miller68,Miller72}, we define the auxiliary functions
$\Phi_{_i}\;(i=B,\;C,\;V)$ through the functions $f_{_i}$ and additional variables
$s_{_i}$, $t_{_i}$, $u_{_i}$ and $v_{_i}$:
\begin{eqnarray}
&&\Phi_{_i}\left(\left.\begin{array}{cc}a_{_i},&b_{_i}\\
c_{_i},&c_{_i}^\prime\end{array}\right|\left.\begin{array}{c}x_{_i},\;y_{_i},\;s_{_i},\;
t_{_i}\\u_{_i},\;v_{_i}\end{array}\right.\right)=
s_{_i}^{a_{_i}}t_{_i}^{b_{_i}}u_{_i}^{c_{_i}-1}v_{_i}^{c_{_i}^\prime-1}
f_{_i}\left(\left.\begin{array}{cc}a_{_i},&b_{_i}\\
c_{_i},&c_{_i}^\prime\end{array}\right|x_{_i},\;y_{_i}\right)\;.
\label{GKZ5}
\end{eqnarray}
Miller's transformation on the functions $f_{_i}$ is to replace the multiplication by
the parameters $a_{_i}$, $b_{_i}$, $c_{_i}$, and $c_{_i}^\prime$ in Eq.~(\ref{GKZ4})
by Euler operators $\hat{\vartheta}_{_{s_{_i}}}$, $\hat{\vartheta}_{_{t_{_i}}}$,
$\hat{\vartheta}_{_{u_{_i}}}$, and $\hat{\vartheta}_{_{v_{_i}}}$
\begin{eqnarray}
&&\hat{\vartheta}_{_{s_{_i}}}\Phi_{_i}\left(\left.\begin{array}{cc}a_{_i},&b_{_i}\\
c_{_i},&c_{_i}^\prime\end{array}\right|\left.\begin{array}{c}x_{_i},\;y_{_i},\;s_{_i},\;
t_{_i}\\u_{_i},\;v_{_i}\end{array}\right.\right)=
a_{_i}\Phi_{_i}\left(\left.\begin{array}{cc}a_{_i},&b_{_i}\\
c_{_i},&c_{_i}^\prime\end{array}\right|\left.\begin{array}{c}x_{_i},\;y_{_i},\;s_{_i},\;
t_{_i}\\u_{_i},\;v_{_i}\end{array}\right.\right)
\;,\nonumber\\
&&\hat{\vartheta}_{_{t_{_i}}}\Phi_{_i}\left(\left.\begin{array}{cc}a_{_i},&b_{_i}\\
c_{_i},&c_{_i}^\prime\end{array}\right|\left.\begin{array}{c}x_{_i},\;y_{_i},\;s_{_i},\;
t_{_i}\\u_{_i},\;v_{_i}\end{array}\right.\right)=
b_{_i}\Phi_{_i}\left(\left.\begin{array}{cc}a_{_i},&b_{_i}\\
c_{_i},&c_{_i}^\prime\end{array}\right|\left.\begin{array}{c}x_{_i},\;y_{_i},\;s_{_i},\;
t_{_i}\\u_{_i},\;v_{_i}\end{array}\right.\right)
\;,\nonumber\\
&&\hat{\vartheta}_{_{u_{_i}}}\Phi_{_i}\left(\left.\begin{array}{cc}a_{_i},&b_{_i}\\
c_{_i},&c_{_i}^\prime\end{array}\right|\left.\begin{array}{c}x_{_i},\;y_{_i},\;s_{_i},\;
t_{_i}\\u_{_i},\;v_{_i}\end{array}\right.\right)=
(c_{_i}-1)\Phi_{_i}\left(\left.\begin{array}{cc}a_{_i},&b_{_i}\\
c_{_i},&c_{_i}^\prime\end{array}\right|\left.\begin{array}{c}x_{_i},\;y_{_i},\;s_{_i},\;
t_{_i}\\u_{_i},\;v_{_i}\end{array}\right.\right)
\;,\nonumber\\
&&\hat{\vartheta}_{_{v_{_i}}}\Phi_{_i}\left(\left.\begin{array}{cc}a_{_i},&b_{_i}\\
c_{_i},&c_{_i}^\prime\end{array}\right|\left.\begin{array}{c}x_{_i},\;y_{_i},\;s_{_i},\;
t_{_i}\\u_{_i},\;v_{_i}\end{array}\right.\right)=
(c_{_i}^\prime-1)\Phi_{_i}\left(\left.\begin{array}{cc}a_{_i},&b_{_i}\\
c_{_i},&c_{_i}^\prime\end{array}\right|\left.\begin{array}{c}x_{_i},\;y_{_i},\;s_{_i},\;
t_{_i}\\u_{_i},\;v_{_i}\end{array}\right.\right)\;,
\label{GKZ6}
\end{eqnarray}
which leads naturally to the notion of GKZ-hypergeometric systems.
In addition, the contiguous relations of Eq.~(\ref{GKZ4}) are rewritten as
\begin{eqnarray}
&&\hat{\cal O}_{_1}^i\Phi_{_i}\left(\left.\begin{array}{cc}a_{_i},&b_{_i}\\
c_{_i},&c_{_i}^\prime\end{array}\right|\left.\begin{array}{c}x_{_i},\;y_{_i},\;s_{_i},\;
t_{_i}\\u_{_i},\;v_{_i}\end{array}\right.\right)=
a_{_i}\Phi_{_i}\left(\left.\begin{array}{cc}a_{_i}+1,&b_{_i}\\
c_{_i},&c_{_i}^\prime\end{array}\right|\left.\begin{array}{c}x_{_i},\;y_{_i},\;s_{_i},\;
t_{_i}\\u_{_i},\;v_{_i}\end{array}\right.\right)
\;,\nonumber\\
&&\hat{\cal O}_{_2}^i\Phi_{_i}\left(\left.\begin{array}{cc}a_{_i},&b_{_i}\\
c_{_i},&c_{_i}^\prime\end{array}\right|\left.\begin{array}{c}x_{_i},\;y_{_i},\;s_{_i},\;
t_{_i}\\u_{_i},\;v_{_i}\end{array}\right.\right)=
b_{_i}\Phi_{_i}\left(\left.\begin{array}{cc}a_{_i},&b_{_i}+1\\
c_{_i},&c_{_i}^\prime\end{array}\right|\left.\begin{array}{c}x_{_i},\;y_{_i},\;s_{_i},\;
t_{_i}\\u_{_i},\;v_{_i}\end{array}\right.\right)
\;,\nonumber\\
&&\hat{\cal O}_{_3}^i\Phi_{_i}\left(\left.\begin{array}{cc}a_{_i},&b_{_i}\\
c_{_i},&c_{_i}^\prime\end{array}\right|\left.\begin{array}{c}x_{_i},\;y_{_i},\;s_{_i},\;
t_{_i}\\u_{_i},\;v_{_i}\end{array}\right.\right)=
(c_{_i}-1)\Phi_{_i}\left(\left.\begin{array}{cc}a_{_i},&b_{_i}\\
c_{_i}-1,&c_{_i}^\prime\end{array}\right|\left.\begin{array}{c}x_{_i},\;y_{_i},\;s_{_i},\;
t_{_i}\\u_{_i},\;v_{_i}\end{array}\right.\right)
\;,\nonumber\\
&&\hat{\cal O}_{_4}^i\Phi_{_i}\left(\left.\begin{array}{cc}a_{_i},&b_{_i}\\
c_{_i},&c_{_i}^\prime\end{array}\right|\left.\begin{array}{c}x_{_i},\;y_{_i},\;s_{_i},\;
t_{_i}\\u_{_i},\;v_{_i}\end{array}\right.\right)=
(c_{_i}^\prime-1)\Phi_{_i}\left(\left.\begin{array}{cc}a_{_i},&b_{_i}\\
c_{_i},&c_{_i}^\prime-1\end{array}\right|\left.\begin{array}{c}x_{_i},\;y_{_i},\;s_{_i},\;
t_{_i}\\u_{_i},\;v_{_i}\end{array}\right.\right)
\;,\nonumber\\
&&\hat{\cal O}_{_5}^i\Phi_{_i}\left(\left.\begin{array}{cc}a_{_i},&b_{_i}\\
c_{_i},&c_{_i}^\prime\end{array}\right|\left.\begin{array}{c}x_{_i},\;y_{_i},\;s_{_i},\;
t_{_i}\\u_{_i},\;v_{_i}\end{array}\right.\right)=
{a_{_i}b_{_i}\over c_{_i}}\Phi_{_i}\left(\left.\begin{array}{cc}a_{_i}+1,&b_{_i}+1\\
c_{_i}+1,&c_{_i}^\prime\end{array}\right|\left.\begin{array}{c}x_{_i},\;y_{_i},\;s_{_i},\;
t_{_i}\\u_{_i},\;v_{_i}\end{array}\right.\right)
\;,\nonumber\\
&&\hat{\cal O}_{_6}^i\Phi_{_i}\left(\left.\begin{array}{cc}a_{_i},&b_{_i}\\
c_{_i},&c_{_i}^\prime\end{array}\right|\left.\begin{array}{c}x_{_i},\;y_{_i},\;s_{_i},\;
t_{_i}\\u_{_i},\;v_{_i}\end{array}\right.\right)=
{a_{_i}b_{_i}\over c_{_i}^\prime}\Phi_{_i}\left(\left.\begin{array}{cc}a_{_i}+1,&b_{_i}+1\\
c_{_i},&c_{_i}^\prime+1\end{array}\right|\left.\begin{array}{c}x_{_i},\;y_{_i},\;s_{_i},\;
t_{_i}\\u_{_i},\;v_{_i}\end{array}\right.\right)\;,
\label{GKZ7}
\end{eqnarray}
where the operators $\hat{\cal O}_{_n}^i,\;(n=1,\cdots,6)$ are
\begin{eqnarray}
&&\hat{\cal O}_{_1}^i=s_{_i}(\hat{\vartheta}_{_{x_{_i}}}+\hat{\vartheta}_{_{y_{_i}}}+\hat{\vartheta}_{_{s_{_i}}})
\;,\nonumber\\
&&\hat{\cal O}_{_2}^i=t_{_i}(\hat{\vartheta}_{_{x_{_i}}}+\hat{\vartheta}_{_{y_{_i}}}+\hat{\vartheta}_{_{t_{_i}}})
\;,\nonumber\\
&&\hat{\cal O}_{_3}^i={1\over u_{_i}}(\hat{\vartheta}_{_{x_{_i}}}+\hat{\vartheta}_{_{u_{_i}}})
\;,\nonumber\\
&&\hat{\cal O}_{_4}^i={1\over v_{_i}}(\hat{\vartheta}_{_{y_{_i}}}+\hat{\vartheta}_{_{v_{_i}}})
\;,\nonumber\\
&&\hat{\cal O}_{_5}^i=s_{_i}t_{_i}u_{_i}\partial_{_{x_{_i}}}
\;,\nonumber\\
&&\hat{\cal O}_{_6}^i=s_{_i}t_{_i}v_{_i}\partial_{_{y_{_i}}}\;.
\label{GKZ8}
\end{eqnarray}
Those operators of Eq.~(\ref{GKZ8}) together with $\hat{\vartheta}_{_{s_{_i}}},\;\hat{\vartheta}_{_{t_{_i}}},\;
\hat{\vartheta}_{_{u_{_i}}},\;\hat{\vartheta}_{_{v_{_i}}}$ define the Lie algebra of the
hypergeometric systems~\cite{Miller68,Miller72}. Under the variable transformation
\begin{eqnarray}
&&z_{_{i,1}}={x_{_i}\over s_{_i}t_{_i}u_{_i}}\;,\;\;z_{_{i,2}}={y_{_i}\over s_{_i}t_{_i}v_{_i}}\;,
\nonumber\\
&&z_{_{i,3}}={1\over s_{_i}}\;,\;\;z_{_{i,4}}={1\over t_{_i}}\;,
\nonumber\\
&&z_{_{i,5}}=u_{_i}\;,\;\;z_{_{i,6}}=v_{_i}\;,
\label{GKZ9}
\end{eqnarray}
the equations in Eq.~(\ref{GKZ6}) are changed as
\begin{eqnarray}
&&\Big(\mathbf{A}\cdot\vec{\vartheta}_{_i}\Big)\Phi_{_i}\left(\left.\begin{array}{cc}a_{_i},&b_{_i}\\
c_{_i},&c_{_i}^\prime\end{array}\right|\left.\begin{array}{c}x_{_i},\;y_{_i},\;s_{_i},\;
t_{_i}\\u_{_i},\;v_{_i}\end{array}\right.\right)=\mathbf{B}\Phi_{_i}\left(\left.\begin{array}{cc}a_{_i},&b_{_i}\\
c_{_i},&c_{_i}^\prime\end{array}\right|\left.\begin{array}{c}x_{_i},\;y_{_i},\;s_{_i},\;
t_{_i}\\u_{_i},\;v_{_i}\end{array}\right.\right)\;,
\label{GKZ10}
\end{eqnarray}
where
\begin{eqnarray}
&&\mathbf{A}=\left(\begin{array}{cccccc}1\;\;&1\;\;&1\;\;&0\;\;&0\;\;&0\;\;\\1\;\;&1\;\;&0\;\;&1\;\;&0\;\;&0\;\;\\
-1\;\;&0\;\;&0\;\;&0\;\;&1\;\;&0\;\;\\0\;\;&-1\;\;&0\;\;&0\;\;&0\;\;&1\;\;\end{array}\right)
\;,\nonumber\\
&&\vec{\vartheta}_{_i}^{\;T}=(\vartheta_{_{z_{_{i,1}}}},\;\vartheta_{_{z_{_{i,2}}}},\;\vartheta_{_{z_{_{i,3}}}},\;
\vartheta_{_{z_{_{i,4}}}},\;\vartheta_{_{z_{_{i,5}}}},\;\vartheta_{_{z_{_{i,6}}}})
\;,\nonumber\\
&&\mathbf{B}^{\;T}=(-a_{_i},\;-b_{_i},\;c_{_i}-1,\;c_{_i}^\prime-1)\;.
\label{GKZ11}
\end{eqnarray}
Correspondingly the universal Gr\"obner basis of the toric ideal associated with $\mathbf{A}$ is
\begin{eqnarray}
&&{\cal U}_{_\mathbf{A}}=\{\partial_{_{z_{_{i,1}}}}\partial_{_{z_{_{i,5}}}}-\partial_{_{z_{_{i,2}}}}\partial_{_{z_{_{i,6}}}},\;
\partial_{_{z_{_{i,1}}}}\partial_{_{z_{_{i,5}}}}-\partial_{_{z_{_{i,3}}}}\partial_{_{z_{_{i,4}}}},\;
\partial_{_{z_{_{i,2}}}}\partial_{_{z_{_{i,6}}}}-\partial_{_{z_{_{i,3}}}}\partial_{_{z_{_{i,4}}}}\}\;.
\label{GKZ12}
\end{eqnarray}
The operators $\mathbf{A}\cdot\vec{\vartheta}_{_i}-\mathbf{B}$ and that from
the set ${\cal U}_{_\mathbf{A}}$ compose the generators of a left ideal~\cite{Cox1991} in the Weyl
algebra $D={\bf C}\langle z_{_{i,1}},\;\cdots,\;z_{_{i,6}},\;\partial_{_{z_{_{i,1}}}},\;\cdots,\;\partial_{_{z_{_{i,6}}}}\rangle$
where ${\bf C}$ denotes the field of complex numbers~\cite{Coutinho1995}. Defining the isomorphism between the commutative
polynomial ring and the Weyl algebra~\cite{M.Saito2000}
\begin{eqnarray}
&&\Psi:\;\;{\bf C}[z_{_{i,1}},\;\cdots,\;z_{_{i,6}},\;\xi_{_{i,1}},\;\cdots,\;\xi_{_{i,6}}]
\rightarrow\;D,\;z_{_i}^\alpha\xi_{_i}^\beta\mapsto z_{_i}^\alpha\partial_{_{z_{_{i}}}}^\beta\;,
\label{GKZ12a}
\end{eqnarray}
one obtains the state polytope~\cite{Sturmfels1995} of the preimage of the universal Gr\"obner basis
in Eq.~(\ref{GKZ12}) as \begin{eqnarray}
&&\xi_{_{i,5}}+\xi_{_{i,6}}\ge1,\;\xi_{_{i,5}}\ge0,\;\xi_{_{i,6}}\ge0,\nonumber\\
&&-\xi_{_{i,5}}\ge-2,\;-\xi_{_{i,6}}\ge-2,\;-\xi_{_{i,5}}-\xi_{_{i,6}}\ge-3
\label{GKZ12b}
\end{eqnarray}
on the hyperplane
\begin{eqnarray}
&&\xi_{_{i,3}}-\xi_{_{i,4}}=0,\;\;\;\xi_{_{i,2}}-\xi_{_{i,6}}=0,\nonumber\\
&&\xi_{_{i,1}}-\xi_{_{i,5}}=0,\;\;\;\xi_{_{i,4}}+\xi_{_{i,5}}+\xi_{_{i,6}}=3\;.
\label{GKZ12c}
\end{eqnarray}
In Eq.~(\ref{GKZ12a}) we take multi-index notation for abbreviation, i.e.
\begin{eqnarray}
&&z_{_{i}}^{\alpha}=\prod\limits_{k=1}^6z_{_{i,k}}^{\alpha_k},\;\;\;
\xi_{_{i}}^{\beta}=\prod\limits_{k=1}^6\xi_{_{i,k}}^{\beta_k}\;,
\label{GKZ12d}
\end{eqnarray}
where $\alpha,\;\beta\in N^6$, and $N=\{0,1,2,\cdots\}$
denotes the set of non-negative integers.
The normal fan of the state polytope in Eq.~(\ref{GKZ12b}) is the Gr\"obner fan
of the corresponding left ideal.
Because codimension$=2$ for all GKZ-hypergeometric systems, the Gr\"obner fan
equals the hypergeometric fan.
These two fans are indispensable in the construction of canonical series solutions
of corresponding GKZ-hypergeometric systems.

\section{Triangulation and construction of series solutions\label{sec3}}
\subsection{Triangulation\label{sec3-1}}
\indent\indent
For an integer $d\times n$-matrix $P\;(d<n)$ of rank $d$ which satisfies the homogeneity
assumption
\begin{eqnarray}
&&(\underbrace{1,\;1,\;\cdots,\;1}\limits_{n})=\sum\limits_{l=1}^dq_{_l}P_{_l}\;,
\label{GKZ12e}
\end{eqnarray}
the Gale transform of $P$~\cite{Gelfand1994} is the $n\times(n-d)$ integer-matrix $Q$ which satisfies
\begin{eqnarray}
&&P\cdot Q=0\;,
\label{GKZ12f}
\end{eqnarray}
where $P_{_l}\;(l=1,\cdots,d)$ is the $l$-th row vector of the integer matrix $P$,
and the combination coefficient $q_{_l}$ is a rational number.
The following matrix $G_{_A}$ is a Gale transform of $A$ in Eq.~(\ref{GKZ11}):
\begin{eqnarray}
&&G_{_A}^T=\left(\begin{array}{cccccc}1\;\;&0\;&-1\;\;&-1\;\;&1\;\;&0\\
0\;\;&1\;&-1\;\;&-1\;\;&0\;\;&1\end{array}\right)\;,
\label{GKZ13}
\end{eqnarray}
whose column vectors compose the secondary fan $\Sigma_{_A}$ of GKZ-hypergeometric system
in Eq.(\ref{GKZ10}). Actually the state polytope in Eq.~(\ref{GKZ12b}) of universal Gr\"obner basis indicates
that the hypergeometric fan ${\cal H}_{_A}$ and the Gr\"obner fan ${\cal G}_{_A}$ all equal
the secondary fan of GKZ-hypergeometric system:
\begin{eqnarray}
&&{\cal H}_{_A}={\cal G}_{_A}=\Sigma_{_A}={\rm Cone}(\{{\bf e}_{_1},{\bf e}_{_2}\})
\bigcup{\rm Cone}(\{{\bf e}_{_1},-{\bf e}_{_1}-{\bf e}_{_2}\})
\bigcup{\rm Cone}(\{{\bf e}_{_2},-{\bf e}_{_1}-{\bf e}_{_2}\})\;,
\label{GKZ14}
\end{eqnarray}
with ${\bf e}_{_1}=(1,\;0)^T$, ${\bf e}_{_2}=(0,\;1)^T$. The cones are defined as~\cite{Cox1998}
\begin{eqnarray}
&&{\rm Cone}(\{{\bf e}_{_1},{\bf e}_{_2}\})=\{\lambda_{_1}{\bf e}_{_1}
+\lambda_{_2}{\bf e}_{_2}\;\big|\;\lambda_{_1},\;\lambda_{_2}\in{\bf R}_{_+}\}
\;,\nonumber\\
&&{\rm Cone}(\{{\bf e}_{_1},-{\bf e}_{_1}-{\bf e}_{_2}\})=\{\lambda_{_1}{\bf e}_{_1}
+\lambda_{_{12}}(-{\bf e}_{_1}-{\bf e}_{_2})\;\big|\;\lambda_{_1},\;\lambda_{_{12}}\in{\bf R}_{_+}\}
\;,\nonumber\\
&&{\rm Cone}(\{{\bf e}_{_2},-{\bf e}_{_1}-{\bf e}_{_2}\})=\{\lambda_{_2}{\bf e}_{_2}
+\lambda_{_{12}}(-{\bf e}_{_1}-{\bf e}_{_2})\;\big|\;\lambda_{_2},\;\lambda_{_{12}}\in{\bf R}_{_+}\}\;,
\label{GKZ15}
\end{eqnarray}
where ${\bf R}_{_+}$ denotes the set of non-negative real numbers.

For a generic weight vector $\omega\in{\rm Cone}(\{{\bf e}_{_1},{\bf e}_{_2}\})$, the corresponding
triangulation~\cite{Sturmfels1995} $\triangle_{_\omega}=\{\sigma_{_1}^a,\;\sigma_{_2}^a,\;
\sigma_{_3}^a,\;\sigma_{_4}^a\}$ is unimodular, and supports the toric ideal
\begin{eqnarray}
&&I=\langle\partial_{_{z_{_{i,1}}}}\partial_{_{z_{_{i,5}}}}-\partial_{_{z_{_{i,3}}}}\partial_{_{z_{_{i,4}}}},
\partial_{_{z_{_{i,2}}}}\partial_{_{z_{_{i,6}}}}-\partial_{_{z_{_{i,3}}}}\partial_{_{z_{_{i,4}}}}\rangle\;,
\label{GKZ16}
\end{eqnarray}
which corresponds to the initial monomial ideal $in_{_\omega}(I)=\langle\partial_{_{z_{_{i,1}}}}\partial_{_{z_{_{i,5}}}},
\partial_{_{z_{_{i,2}}}}\partial_{_{z_{_{i,6}}}}\rangle$ with $i=B,\;V,\;C$.
Each facet $\sigma_{_j}^a,\;(j=1,\;2,\;3,\;4)$ of the simplicial complex $\triangle_{_\omega}$ is
the index set of an invertible $4\times4$-submatrix $A_{_{\sigma_{_j}^a}}$ of $A$:
\begin{eqnarray}
&&\sigma_{_1}^a=\{1,\;2,\;3,\;4\}\;,\;\;\sigma_{_2}^a=\{1,\;3,\;4,\;6\}\;,
\nonumber\\
&&\sigma_{_3}^a=\{2,\;3,\;4,\;5\}\;,\;\;\sigma_{_4}^a=\{3,\;4,\;5,\;6\}\;.
\label{GKZ17}
\end{eqnarray}
Certainly four standard pairs~\cite{M.Saito2000} $(1,\;\sigma_{_j}^a),\;(j=1,\;2,\;3,\;4)$
produce the following exponent vectors of initial monomials of series solutions
\begin{eqnarray}
&&p_{_{\sigma_{_1}^a}}=(1-c_{_i},\;1-c_{_i}^\prime,\;c_{_i}+c_{_i}^\prime-a_{_i}-2,\;c_{_i}+c_{_i}^\prime-b_{_i}-2,\;0,\;0)
\;,\nonumber\\
&&p_{_{\sigma_{_2}^a}}=(c_{_i}-b_{_i}-1,\;0,\;b_{_i}-a_{_i}-c_{_i}+1,\;1-c_{_i},\;0,\;c_{_i}^\prime-1)
\;,\nonumber\\
&&p_{_{\sigma_{_3}^a}}=(0,\;c_{_i}^\prime-1,\;1-a_{_i}-c_{_i}^\prime,\;1-b_{_i}-c_{_i}^\prime,\;c_{_i}-1,\;0)
\;,\nonumber\\
&&p_{_{\sigma_{_4}^a}}=(0,\;0,\;-a_{_i},\;-b_{_i},\;c_{_i}-1,\;c_{_i}^\prime-1)\;.
\label{GKZ18}
\end{eqnarray}

For a generic weight vector
$\omega\in{\rm Cone}(\{{\bf e}_{_2},-{\bf e}_{_1}-{\bf e}_{_2}\})$, the corresponding
triangulation $\triangle_{_\omega}=\{\sigma_{_1}^b,\;\sigma_{_2}^b,\;
\sigma_{_3}^b,\;\sigma_{_4}^b\}$ is also unimodular, and supports similarly the toric ideal
\begin{eqnarray}
&&I=\langle\partial_{_{z_{_{i,2}}}}\partial_{_{z_{_{i,6}}}}-\partial_{_{z_{_{i,1}}}}\partial_{_{z_{_{i,5}}}},
\partial_{_{z_{_{i,3}}}}\partial_{_{z_{_{i,4}}}}-\partial_{_{z_{_{i,1}}}}\partial_{_{z_{_{i,5}}}}\rangle\;,
\label{GKZ19}
\end{eqnarray}
which corresponds to the initial monomial ideal $in_{_\omega}(I)=\langle\partial_{_{z_{_{i,2}}}}\partial_{_{z_{_{i,6}}}},
\partial_{_{z_{_{i,3}}}}\partial_{_{z_{_{i,4}}}}\rangle$.
Each facet $\sigma_{_j}^b,\;(j=1,\;2,\;3,\;4)$ of the simplicial complex $\triangle_{_\omega}$ is
the index set of an invertible $4\times4$-submatrix $A_{_{\sigma_{_i}^b}}$ of $A$:
\begin{eqnarray}
&&\sigma_{_1}^b=\{1,\;2,\;3,\;5\}\;,\;\;\sigma_{_2}^b=\{1,\;2,\;4,\;5\}\;,
\nonumber\\
&&\sigma_{_3}^b=\{1,\;3,\;5,\;6\}\;,\;\;\sigma_{_4}^b=\{1,\;4,\;5,\;6\}\;.
\label{GKZ20}
\end{eqnarray}
Correspondingly four standard pairs $(1,\;\sigma_{_j}^b),\;(j=1,\;2,\;3,\;4)$
induce the following exponent vectors of initial monomials for series solutions
\begin{eqnarray}
&&p_{_{\sigma_{_1}^b}}=(2-c_{_i}-c_{_i}^\prime,\;c_{_i}+c_{_i}^\prime-b_{_i}-2,\;b_{_i}-a_{_i}^\prime,\;0,\;1-c_{_i}^\prime,\;0)
\;,\nonumber\\
&&p_{_{\sigma_{_2}^b}}=(c_{_i}^\prime-a_{_i}-1,\;1-c_{_i}^\prime,\;0,\;a_{_i}-b_{_i},\;c_{_i}+c_{_i}^\prime-a_{_i}-2,\;0)
\;,\nonumber\\
&&p_{_{\sigma_{_3}^b}}=(-b_{_i},\;0,\;b_{_i}-a_{_i},\;0,\;c_{_i}-b_{_i}-1,\;c_{_i}^\prime-1)
\;,\nonumber\\
&&p_{_{\sigma_{_4}^b}}=(-a_{_i},\;0,\;0,\;a_{_i}-b_{_i},\;c_{_i}-a_{_i}-1,\;c_{_i}^\prime-1)\;.
\label{GKZ21}
\end{eqnarray}

Finally for a generic weight vector
$\omega\in{\rm Cone}(\{{\bf e}_{_1},-{\bf e}_{_1}-{\bf e}_{_2}\})$, the corresponding
triangulation $\triangle_{_\omega}=\{\sigma_{_1}^c,\;\sigma_{_2}^c,\;
\sigma_{_3}^c,\;\sigma_{_4}^c\}$ is unimodular, and supports the toric ideal
\begin{eqnarray}
&&I=\langle\partial_{_{z_{_{i,1}}}}\partial_{_{z_{_{i,5}}}}-\partial_{_{z_{_{i,2}}}}\partial_{_{z_{_{i,6}}}},
\partial_{_{z_{_{i,3}}}}\partial_{_{z_{_{i,4}}}}-\partial_{_{z_{_{i,2}}}}\partial_{_{z_{_{i,6}}}}\rangle\;,
\label{GKZ22}
\end{eqnarray}
which corresponds to the initial monomial ideal $in_{_\omega}(I)=\langle\partial_{_{z_{_{i,1}}}}\partial_{_{z_{_{i,5}}}},
\partial_{_{z_{_{i,3}}}}\partial_{_{z_{_{i,4}}}}\rangle$.
Each facet $\sigma_{_j}^c,\;(j=1,\;2,\;3,\;4)$ of the simplicial complex $\triangle_{_\omega}$ is
the index set of an invertible $4\times4$-submatrix $A_{_{\sigma_{_j}^c}}$ of matrix $A$:
\begin{eqnarray}
&&\sigma_{_1}^c=\{1,\;2,\;3,\;6\}\;,\;\;\sigma_{_2}^c=\{1,\;2,\;4,\;6\}\;,
\nonumber\\
&&\sigma_{_3}^c=\{2,\;3,\;5,\;6\}\;,\;\;\sigma_{_4}^c=\{2,\;4,\;5,\;6\}\;.
\label{GKZ23}
\end{eqnarray}
Correspondingly four standard pairs $(1,\;\sigma_{_j}^c),\;(j=1,\;2,\;3,\;4)$
give the following exponent vectors of initial monomials for series solutions
\begin{eqnarray}
&&p_{_{\sigma_{_1}^c}}=(1-c_{_i},\;c_{_i}-b_{_i}-1,\;a_{_i}-b_{_i},\;0,\;0,\;c_{_i}+c_{_i}^\prime-b_{_i}-2)
\;,\nonumber\\
&&p_{_{\sigma_{_2}^c}}=(1-c_{_i},\;c_{_i}-a_{_i}-1,\;0,\;a_{_i}-b_{_i},\;0,\;c_{_i}+c_{_i}^\prime-a_{_i}-2)
\;,\nonumber\\
&&p_{_{\sigma_{_3}^c}}=(0,\;-b_{_i},\;b_{_i}-a_{_i},\;0,\;c_{_i}-1,\;c_{_i}^\prime-b_{_i}-1)
\;,\nonumber\\
&&p_{_{\sigma_{_4}^c}}=(0,\;-a_{_i},\;0,\;a_{_i}-b_{_i},\;c_{_i}-1,\;c_{_i}^\prime-a_{_i}-1)\;.
\label{GKZ24}
\end{eqnarray}
\begin{figure}[h]
\setlength{\unitlength}{1cm}
\centering
\vspace{0.0cm}\hspace{-1.5cm}
\includegraphics[height=8cm,width=8.0cm]{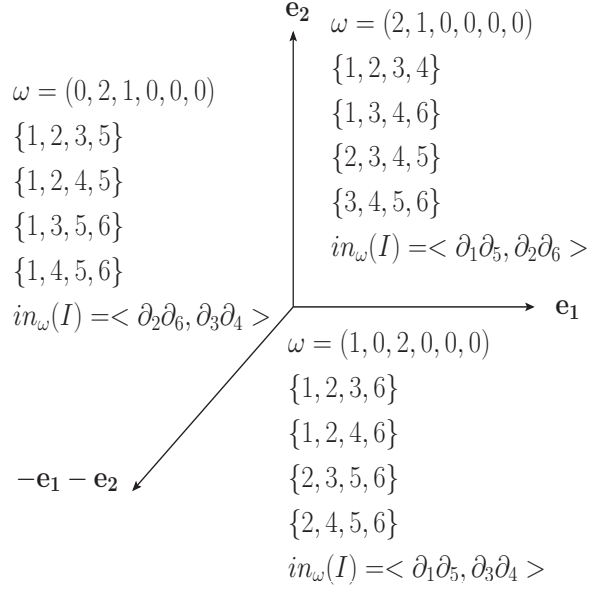}
\vspace{0cm}
\caption[]{The secondary fan of GKZ-hypergeometric system
in Eq.(\ref{GKZ10}), $\omega$ in each cone is a representative weight vector.}
\label{fig1}
\end{figure}
\subsection{Construction of canonical series solutions\label{sec3-2}}
\indent\indent
The integer kernel of the matrix $A$ is defined as
\begin{eqnarray}
&&{\rm ker}_{\bf Z}(A)=\{u\in{\bf Z}^6:\;\;A\cdot u=0\}
\nonumber\\
&&\hspace{1.5cm}=
\{N(1,\;0,\;-1,\;-1,\;1,\;0),\;N(-1,\;0,\;1,\;1,\;-1,\;0),
\nonumber\\
&&\hspace{2.2cm}
N(0,\;1,\;-1,\;-1,\;0,\;1),\;N(0,\;-1,\;1,\;1,\;0,\;-1),
\nonumber\\
&&\hspace{2.2cm}
N(-1,\;1,\;0,\;0,\;-1,\;1),\;N(1,\;-1,\;0,\;0,\;1,\;-1)\}\;.
\label{GKZ25}
\end{eqnarray}
The vector $u\in{\rm ker}_{\bf Z}(A)$ can be decomposed into positive and negative
part, $u=u_+-u_-$, where $u_+$ and $u_-$ are non-negative vectors with disjoint supports.
In order to construct canonical series solutions of GKZ-hypergeometric system,
we define the negative support of any vector $v=(v_{_1},\;\cdots,\;v_{_n})\in{\bf R}^n$ as
\begin{eqnarray}
&&nsupp(v)=\{i\in\{1,\;2,\;\cdots,\;n\}\;:\;v_{_i}\;{\rm is\;a\;negative\;integer}\}\;.
\label{GKZ26}
\end{eqnarray}
Furthermore we introduce the following subset of ${\rm ker}_{\bf Z}(A)$
\begin{eqnarray}
&&N_p=\{u\in{\rm ker}_{\bf Z}(A)\;:\;nsupp(p)=nsupp(p+u)\}\;.
\label{GKZ27}
\end{eqnarray}
With an exponent vector $p$ of the initial monomial, the corresponding canonical series
solution of the hypergeometric system Eq.~(\ref{GKZ10}) is well-defined:
\begin{eqnarray}
&&\phi_p=\sum\limits_{u\in N_p}{[p]_{u_-}\over[p+u]_{u_+}}z_{_i}^{p+u}\;,
\label{GKZ28}
\end{eqnarray}
where the abbreviations
\begin{eqnarray}
&&z_{_i}^{p}=\prod\limits_{j=1}^6z_{_{i,j}}^{p_j}
\;,\nonumber\\
&&[p]_{u_-}=\prod\limits_{k:u_{_k}<0}\prod\limits_{j=1}^{-u_{_k}}(p_k-j+1)
\;,\nonumber\\
&&[p+u]_{u_+}=\prod\limits_{k:u_{_k}>0}\prod\limits_{j=1}^{u_{_k}}(p_k+j)\;.
\label{GKZ29}
\end{eqnarray}

For one-loop self energy and two-loop vacuum, $p_{_{\sigma_{_1}^a}}=(D/2-1,\;D/2-1,\;-D/2,\;-1,0,\;0)$,
where $p_{_{\sigma_{_1}^a,4}}=-1$ is a negative integer. To construct canonical series solution properly,
we perturb the vector ${\bf B}$ in the direction of ${\bf B}^\prime=(0,\;1,\;0,\;0)^T$.
Choosing the perturbed parameter vector ${\bf B}+\epsilon{\bf B}^\prime$, we modify
the exponent vector as $p_{_{\sigma_{_1}^a}}^{\;\prime}=(D/2-1,\;D/2-1,\;-D/2,\;-1+\epsilon,0,\;0)$.
Correspondingly the set
\begin{eqnarray}
&&N_{p_{_{\sigma_{_1}^a}}^{\;\prime}}=\{N(1,\;0,\;-1,\;-1,\;1,\;0)\;,
N(0,\;1,\;-1,\;-1,\;0,\;1)
\;,\nonumber\\
&&\hspace{1.6cm}N(1,\;0,\;-1,\;-1,\;1,\;0)+N(0,\;1,\;-1,\;-1,\;0,\;1)\}\;.
\label{GKZ30}
\end{eqnarray}
Using Eq.~(\ref{GKZ28}), one derives
\begin{eqnarray}
&&\phi_{p_{_{\sigma_{_1}^a}}^{\;\prime}}={(z_{_{i,1}}z_{_{i,2}})^{D/2-1}\over
z_{_{i,3}}^{D/2}z_{_{i,4}}^{1-\epsilon}}
\Big\{1+\sum\limits_{n_{_1}=1}^\infty\prod\limits_{j=1}^{n_{_1}}
{(1-{D\over2}-j)(\epsilon-j)\over j({D\over2}-1+j)}\Big({z_{_{i,1}}z_{_{i,5}}\over
z_{_{i,3}}z_{_{i,4}}}\Big)^{n_{_1}}
\nonumber\\
&&\hspace{1.5cm}
+\sum\limits_{n_{_2}=1}^\infty\prod\limits_{j=1}^{n_{_2}}
{(1-{D\over2}-j)(\epsilon-j)\over j({D\over2}-1+j)}\Big({z_{_{i,2}}z_{_{i,6}}\over
z_{_{i,3}}z_{_{i,4}}}\Big)^{n_{_2}}
\nonumber\\
&&\hspace{1.5cm}
+\sum\limits_{n_{_1}=1}^\infty\sum\limits_{n_{_2}=1}^\infty
{\prod\limits_{j=1}^{n_{_1}+n_{_2}}(1-{D\over2}-j)(\epsilon-j)\over
\prod\limits_{j_{_1}=1}^{n_{_1}}j_{_1}({D\over2}-1+j_{_1})\prod\limits_{j_{_2}=1}^{n_{_2}}
j_{_2}({D\over2}-1+j_{_2})}\Big({z_{_{i,1}}z_{_{i,5}}\over
z_{_{i,3}}z_{_{i,4}}}\Big)^{n_{_1}}\Big({z_{_{i,2}}z_{_{i,6}}\over
z_{_{i,3}}z_{_{i,4}}}\Big)^{n_{_2}}\Big\}
\nonumber\\
&&\hspace{1.2cm}=
{(z_{_{i,1}}z_{_{i,2}})^{D/2-1}\over
z_{_{i,3}}^{D/2}z_{_{i,4}}^{1-\epsilon}}\Big\{1+\sum\limits_{n_{_1}=1}^\infty
{({D\over2})_{_{n_{_1}}}(1-\epsilon)_{_{n_{_1}}}\over n_{_1}!({D\over2})_{_{n_{_1}}}}\Big({z_{_{i,1}}z_{_{i,5}}\over
z_{_{i,3}}z_{_{i,4}}}\Big)^{n_{_1}}
\nonumber\\
&&\hspace{1.5cm}
+\sum\limits_{n_{_2}=1}^\infty{({D\over2})_{_{n_{_2}}}(1-\epsilon)_{_{n_{_2}}}
\over n_{_2}!({D\over2})_{_{n_{_2}}}}\Big({z_{_{i,2}}z_{_{i,6}}\over
z_{_{i,3}}z_{_{i,4}}}\Big)^{n_{_2}}
\nonumber\\
&&\hspace{1.5cm}
+\sum\limits_{n_{_1}=1}^\infty\sum\limits_{n_{_2}=1}^\infty
{({D\over2})_{_{n_{_1}+n_{_2}}}(1-\epsilon)_{_{n_{_1}+n_{_2}}}\over
n_{_1}!n_{_2}!({D\over2})_{_{n_{_1}}}({D\over2})_{_{n_{_2}}}}\Big({z_{_{i,1}}z_{_{i,5}}\over
z_{_{i,3}}z_{_{i,4}}}\Big)^{n_{_1}}\Big({z_{_{i,2}}z_{_{i,6}}\over
z_{_{i,3}}z_{_{i,4}}}\Big)^{n_{_2}}\Big\}\;.
\label{GKZ31}
\end{eqnarray}
Then
\begin{eqnarray}
&&\phi_{p_{_{\sigma_{_1}^a}}}=\lim\limits_{\epsilon\rightarrow0}\phi_{p_{_{\sigma_{_1}^a}}^{\;\prime}}
\nonumber\\
&&\hspace{0.9cm}=
{(z_{_{i,1}}z_{_{i,2}})^{D/2-1}\over z_{_{i,3}}^{D/2}z_{_{i,4}}}F_{_4}
\left(\left.\begin{array}{cc}1,&{D\over2}\\
{D\over2},&{D\over2}\end{array}\right|{z_{_{i,1}}z_{_{i,5}}\over
z_{_{i,3}}z_{_{i,4}}},\;{z_{_{i,2}}z_{_{i,6}}\over
z_{_{i,3}}z_{_{i,4}}}\right)\;,
\label{GKZ32}
\end{eqnarray}
where $F_{_4}$ denotes the fourth Appell function~\cite{L.J.Slater66}.
Similarly we have
\begin{eqnarray}
&&\phi_{p_{_{\sigma_{_2}^a}}}=z_{_{i,1}}^{D/2-1}z_{_{i,3}}^{-1}z_{_{i,4}}^{D/2-2}z_{_{i,6}}^{1-D/2}F_{_4}
\left(\left.\begin{array}{cc}1,&2-{D\over2}\\
{D\over2},&2-{D\over2}\end{array}\right|{z_{_{i,1}}z_{_{i,5}}\over
z_{_{i,3}}z_{_{i,4}}},\;{z_{_{i,2}}z_{_{i,6}}\over
z_{_{i,3}}z_{_{i,4}}}\right)
\;,\nonumber\\
&&\phi_{p_{_{\sigma_{_3}^a}}}=z_{_{i,2}}^{D/2-1}z_{_{i,3}}^{-1}z_{_{i,4}}^{D/2-2}z_{_{i,5}}^{1-D/2}F_{_4}
\left(\left.\begin{array}{cc}1,&2-{D\over2}\\
2-{D\over2},&{D\over2}\end{array}\right|{z_{_{i,1}}z_{_{i,5}}\over
z_{_{i,3}}z_{_{i,4}}},\;{z_{_{i,2}}z_{_{i,6}}\over
z_{_{i,3}}z_{_{i,4}}}\right)
\;,\nonumber\\
&&\phi_{p_{_{\sigma_{_4}^a}}}=z_{_{i,3}}^{D/2-2}z_{_{i,4}}^{D-3}z_{_{i,5}}^{1-D/2}z_{_{i,6}}^{1-D/2}F_{_4}
\left(\left.\begin{array}{cc}3-D,&2-{D\over2}\\
2-{D\over2},&2-{D\over2}\end{array}\right|{z_{_{i,1}}z_{_{i,5}}\over
z_{_{i,3}}z_{_{i,4}}},\;{z_{_{i,2}}z_{_{i,6}}\over
z_{_{i,3}}z_{_{i,4}}}\right)
\;,\nonumber\\
&&\phi_{p_{_{\sigma_{_1}^b}}}=z_{_{i,1}}^{D/2-2}z_{_{i,2}}^{D/2-1}z_{_{i,3}}^{1-D/2}z_{_{i,5}}^{-1}F_{_4}
\left(\left.\begin{array}{cc}1,&2-{D\over2}\\
2-{D\over2},&{D\over2}\end{array}\right|{z_{_{i,3}}z_{_{i,4}}\over
z_{_{i,1}}z_{_{i,5}}},\;{z_{_{i,2}}z_{_{i,6}}\over
z_{_{i,1}}z_{_{i,5}}}\right)
\;,\nonumber\\
&&\phi_{p_{_{\sigma_{_2}^b}}}=z_{_{i,1}}^{-1}z_{_{i,2}}^{D/2-1}z_{_{i,4}}^{D/2-1}z_{_{i,5}}^{-D/2}F_{_4}
\left(\left.\begin{array}{cc}1,&{D\over2}\\
{D\over2},&{D\over2}\end{array}\right|{z_{_{i,3}}z_{_{i,4}}\over
z_{_{i,1}}z_{_{i,5}}},\;{z_{_{i,2}}z_{_{i,6}}\over
z_{_{i,1}}z_{_{i,5}}}\right)
\;,\nonumber\\
&&\phi_{p_{_{\sigma_{_3}^b}}}=z_{_{i,1}}^{D-3}z_{_{i,3}}^{1-D/2}z_{_{i,5}}^{D/2-2}z_{_{i,6}}^{1-D/2}F_{_4}
\left(\left.\begin{array}{cc}3-D,&2-{D\over2}\\
2-{D\over2},&2-{D\over2}\end{array}\right|{z_{_{i,3}}z_{_{i,4}}\over
z_{_{i,1}}z_{_{i,5}}},\;{z_{_{i,2}}z_{_{i,6}}\over
z_{_{i,1}}z_{_{i,5}}}\right)
\;,\nonumber\\
&&\phi_{p_{_{\sigma_{_4}^b}}}=z_{_{i,1}}^{D/2-2}z_{_{i,4}}^{D/2-1}z_{_{i,5}}^{-1}z_{_{i,6}}^{1-D/2}F_{_4}
\left(\left.\begin{array}{cc}1,&2-{D\over2}\\
{D\over2},&2-{D\over2}\end{array}\right|{z_{_{i,3}}z_{_{i,4}}\over
z_{_{i,1}}z_{_{i,5}}},\;{z_{_{i,2}}z_{_{i,6}}\over
z_{_{i,1}}z_{_{i,5}}}\right)
\;,\nonumber\\
&&\phi_{p_{_{\sigma_{_1}^c}}}=z_{_{i,1}}^{D/2-1}z_{_{i,2}}^{D/2-2}z_{_{i,3}}^{1-D/2}z_{_{i,6}}^{-1}F_{_4}
\left(\left.\begin{array}{cc}1,&2-{D\over2}\\
{D\over2},&2-{D\over2}\end{array}\right|{z_{_{i,1}}z_{_{i,5}}\over
z_{_{i,2}}z_{_{i,6}}},\;{z_{_{i,3}}z_{_{i,4}}\over
z_{_{i,2}}z_{_{i,6}}}\right)
\;,\nonumber\\
&&\phi_{p_{_{\sigma_{_2}^c}}}=z_{_{i,1}}^{D/2-1}z_{_{i,2}}^{-1}z_{_{i,4}}^{D/2-1}z_{_{i,6}}^{-D/2}F_{_4}
\left(\left.\begin{array}{cc}1,&{D\over2}\\
{D\over2},&{D\over2}\end{array}\right|{z_{_{i,1}}z_{_{i,5}}\over
z_{_{i,2}}z_{_{i,6}}},\;{z_{_{i,3}}z_{_{i,4}}\over
z_{_{i,2}}z_{_{i,6}}}\right)
\;,\nonumber\\
&&\phi_{p_{_{\sigma_{_3}^c}}}=z_{_{i,2}}^{D-3}z_{_{i,3}}^{1-D/2}z_{_{i,5}}^{1-D/2}z_{_{i,6}}^{D/2-2}F_{_4}
\left(\left.\begin{array}{cc}3-D,&2-{D\over2}\\
2-{D\over2},&2-{D\over2}\end{array}\right|{z_{_{i,1}}z_{_{i,5}}\over
z_{_{i,2}}z_{_{i,6}}},\;{z_{_{i,3}}z_{_{i,4}}\over
z_{_{i,2}}z_{_{i,6}}}\right)
\;,\nonumber\\
&&\phi_{p_{_{\sigma_{_4}^c}}}=z_{_{i,2}}^{D/2-2}z_{_{i,4}}^{D/2-1}z_{_{i,5}}^{1-D/2}z_{_{i,6}}^{-1}F_{_4}
\left(\left.\begin{array}{cc}1,&2-{D\over2}\\
2-{D\over2},&{D\over2}\end{array}\right|{z_{_{i,1}}z_{_{i,5}}\over
z_{_{i,2}}z_{_{i,6}}},\;{z_{_{i,3}}z_{_{i,4}}\over
z_{_{i,2}}z_{_{i,6}}}\right)\;.
\label{GKZ33}
\end{eqnarray}
For the fourth Appell function
\begin{eqnarray}
&&F_{_4}\left(\left.\begin{array}{cc}a,&b\\c,&c^\prime\end{array}\right|x,\;y\right)
=\sum\limits_{n_{_1}=0}^\infty\sum_{n_{_2}=0}^\infty{(a)_{_{n_{_1}}+n_{_2}}(b)_{_{n_{_1}}+n_{_2}}\over
n_{_1}!n_{_2}!(c)_{_{n_{_1}}}(c^\prime)_{_{n_{_1}}}}x^{n_{_1}}y^{n_{_2}}
\nonumber\\
&&\hspace{3.2cm}=
\sum\limits_{n_{_1}=0}^\infty\sum_{n_{_2}=0}^\infty A_{_{n_{_1},n_{_2}}}x^{n_{_1}}y^{n_{_2}}\;,
\label{GKZ33a}
\end{eqnarray}
the adjacent ratios of the coefficients are
\begin{eqnarray}
&&\Phi_{_1}^\prime(n_{_1},n_{_2})={A_{_{1+n_{_1},n_{_2}}}\over A_{_{n_{_1},n_{_2}}}}
={(1+n_{_1}+n_{_2})(3-D/2+n_{_1}+n_{_2})\over(1+n_{_1})(3-D/2+n_{_1})}
\;,\nonumber\\
&&\Phi_{_2}^\prime(n_{_1},n_{_2})={A_{_{n_{_1},1+n_{_2}}}\over A_{_{n_{_1},n_{_2}}}}
={(1+n_{_1}+n_{_2})(3-D/2+n_{_1}+n_{_2})\over(1+n_{_2})(3-D/2+n_{_2})}\;.
\label{GKZ33b}
\end{eqnarray}
To investigate the absolutely and uniformly convergent
region of the double series in Eq.(\ref{GKZ33a}), one takes
\begin{eqnarray}
&&r_{_x}=t_{_x}^2=\lim\limits_{\lambda\rightarrow\infty}
{1\over\Phi_{_1}^\prime(\lambda n_{_1},\lambda n_{_2})}
={n_{_1}^2\over(n_{_1}+n_{_2})^2}={1\over(1+t)^2}
\;,\nonumber\\
&&r_{_y}=t_{_y}^2=\lim\limits_{\lambda\rightarrow\infty}
{1\over\Phi_{_2}^\prime(\lambda n_{_1},\lambda n_{_2})}
={n_{_2}^2\over(n_{_1}+n_{_2})^2}={t^2\over(1+t)^2}\;,
\label{GKZ33c}
\end{eqnarray}
with $t=n_{_2}/n_{_1},\;r_{_x}=|x|,\;r_{_y}=|y|$, respectively.
The generator of the principal ideal $\langle t_{_x}^2(1+t)^2-1,\;
t_{_y}^2(1+t)^2-t^2\rangle\bigcap{\bf C}[t_{_x},t_{_y}]$ is
\begin{eqnarray}
&&g(t_{_x},\;t_{_y})=1-2(t_{_x}^2+t_{_y}^2)+(t_{_x}^2-t_{_y}^2)^2
\nonumber\\
&&\hspace{1.6cm}=
(-1+t_{_x}-t_{_y})(1+t_{_x}-t_{_y})(-1+t_{_x}+t_{_y})(1+t_{_x}+t_{_y})\;,
\label{GKZ33d}
\end{eqnarray}
where ${\bf C}[t_{_x},t_{_y}]$ denotes the polynomial ring of $t_{_x},\;t_{_y}$
on the field ${\bf C}$. Since $t_{_x},\;t_{_y}\ge0$ the equation
$g(t_{_x},\;t_{_y})=0$ gives the Cartesian curve of the double power series in Eq.(\ref{GKZ33a}) as
\begin{eqnarray}
&&\sqrt{|x|}+\sqrt{|y|}=1\;.
\label{GKZ33e}
\end{eqnarray}
For convenience we denote the region
surrounded by the coordinate axes and the Cartesian curve in the
positive quadrant of the plane $Or_{_x}r_{_y}$ by $C$, and denote
the rectangle by $D$ in the positive quadrant of the plane $Or_{_x}r_{_y}$
bounded by the coordinate axes and the straight lines parallel
to the coordinate axes $r_{_x}=1$, and $r_{_y}=1$.
According to Horn's study of convergence of the hypergeometric series \cite{Horn1889},
one finds the well-known conclusion \cite{L.J.Slater66} that
the double power series in Eq.(\ref{GKZ33a}) absolutely and uniformly converges
in the intersection of the regions $C$ and $D$ in the plane $Or_{_x}r_{_y}$,
i.e. $\sqrt{|x|}+\sqrt{|y|}<1$.

Finally we set $z_{_{i,1}}=x_{_i},\;z_{_{i,2}}=y_{_i}$, $z_{_{i,k}}=1$ with
$i=B,\;V$ and $k=3,\cdots,6$, and formulate the Feynman integrals as the linear combinations
of canonical series solutions in the corresponding parameter space.
\begin{itemize}
\item For $|x_{_i}|\le1,\;|y_{_i}|\le1$, the Feynman integral is
\begin{eqnarray}
&&S_{_{a,i}}(x_{_i},y_{_i})=A_{_{a,i}}(x_{_{i}}y_{_{i}})^{D/2-1}F_{_4}
\left(\left.\begin{array}{cc}1,&{D\over2}\\
{D\over2},&{D\over2}\end{array}\right|x_{_{i}},\;y_{_{i}}\right)
\nonumber\\
&&\hspace{2.2cm}
+B_{_{a,i}}x_{_{i}}^{D/2-1}F_{_4}
\left(\left.\begin{array}{cc}1,&2-{D\over2}\\
{D\over2},&2-{D\over2}\end{array}\right|x_{_{i}},\;y_{_{i}}\right)
\nonumber\\
&&\hspace{2.2cm}
+C_{_{a,i}}y_{_{i}}^{D/2-1}F_{_4}\left(\left.\begin{array}{cc}1,&2-{D\over2}\\
2-{D\over2},&{D\over2}\end{array}\right|x_{_{i}},\;y_{_{i}}\right)
\nonumber\\
&&\hspace{2.2cm}
+D_{_{a,i}}F_{_4}\left(\left.\begin{array}{cc}3-D,&2-{D\over2}\\
2-{D\over2},&2-{D\over2}\end{array}\right|x_{_{i}},\;y_{_{i}}\right)\;,
\label{GKZ33f}
\end{eqnarray}
which is convergent in the region $\sqrt{|x_{_{i}}|}+\sqrt{|y_{_{i}}|}<1$.

\item For $|x_{_i}|\ge1,\;|y_{_i}|\le1$, the Feynman integral is
\begin{eqnarray}
&&S_{_{b,i}}(x_{_i},y_{_i})=A_{_{b,i}}x_{_{i}}^{D/2-2}y_{_{i}}^{D/2-1}F_{_4}
\left(\left.\begin{array}{cc}1,&2-{D\over2}\\
2-{D\over2},&{D\over2}\end{array}\right|{1\over x_{_{i}}},\;{y_{_{i}}\over x_{_i}}\right)
\nonumber\\
&&\hspace{2.2cm}
+B_{_{b,i}}x_{_{i}}^{-1}y_{_{i}}^{D/2-1}F_{_4}\left(\left.\begin{array}{cc}1,&{D\over2}\\
{D\over2},&{D\over2}\end{array}\right|{1\over x_{_{i}}},\;{y_{_{i}}\over x_{_i}}\right)
\nonumber\\
&&\hspace{2.2cm}
+C_{_{b,i}}x_{_{i}}^{D-3}F_{_4}\left(\left.\begin{array}{cc}3-D,&2-{D\over2}\\
2-{D\over2},&2-{D\over2}\end{array}\right|{1\over x_{_{i}}},\;{y_{_{i}}\over x_{_i}}\right)
\nonumber\\
&&\hspace{2.2cm}
+D_{_{b,i}}x_{_{i}}^{D/2-2}F_{_4}\left(\left.\begin{array}{cc}1,&2-{D\over2}\\
{D\over2},&2-{D\over2}\end{array}\right|{1\over x_{_{i}}},\;{y_{_{i}}\over x_{_i}}\right)\;,
\label{GKZ33g}
\end{eqnarray}
which is convergent in the region $1+\sqrt{|y_{_{i}}|}<\sqrt{|x_{_{i}}|}$.

\item For $|x_{_i}|\le1,\;|y_{_i}|\ge1$, the Feynman integral is
\begin{eqnarray}
&&S_{_{c,i}}(x_{_i},y_{_i})=A_{_{c,i}}x_{_{i}}^{D/2-1}y_{_{i}}^{D/2-2}F_{_4}
\left(\left.\begin{array}{cc}1,&2-{D\over2}\\{D\over2},&2-{D\over2}\end{array}\right|
{x_{_{i}}\over y_{_{i}}},\;{1\over y_{_{i}}}\right)
\nonumber\\
&&\hspace{2.2cm}
+B_{_{c,i}}x_{_{i}}^{D/2-1}y_{_{i}}^{-1}F_{_4}
\left(\left.\begin{array}{cc}1,&{D\over2}\\{D\over2},&{D\over2}\end{array}\right|
{x_{_{i}}\over y_{_{i}}},\;{1\over y_{_{i}}}\right)
\nonumber\\
&&\hspace{2.2cm}
+C_{_{c,i}}y_{_{i}}^{D-3}F_{_4}
\left(\left.\begin{array}{cc}3-D,&2-{D\over2}\\2-{D\over2},&2-{D\over2}\end{array}\right|
{x_{_{i}}\over y_{_{i}}},\;{1\over y_{_{i}}}\right)
\nonumber\\
&&\hspace{2.2cm}
+D_{_{c,i}}y_{_{i}}^{D/2-2}F_{_4}
\left(\left.\begin{array}{cc}1,&2-{D\over2}\\2-{D\over2},&{D\over2}\end{array}\right|
{x_{_{i}}\over y_{_{i}}},\;{1\over y_{_{i}}}\right)\;,
\label{GKZ33h}
\end{eqnarray}
which is convergent in the region $1+\sqrt{|x_{_{i}}|}<\sqrt{|y_{_{i}}|}$.
\end{itemize}
In order to determine those integration constants, i.e. the combination coefficients
$A_{_{\sigma,i}}$, $B_{_{\sigma,i}}$, $C_{_{\sigma,i}}$, $D_{_{\sigma,i}}$ with
$\sigma=a,\;b,\;c$ and $i=B,\;V$, we utilize expressions
of the Feynman integrals at some special points of the parameter space. For the Feynman integral
of one-loop self energy diagram $B_{_0}(p^2,\;m_{_1}^2,\;m_{_2}^2)$, we employ the following
expressions
\begin{eqnarray}
&&B_{_0}(\Lambda^2,\;0,\;0)={i\Gamma(2-{D\over2})\Gamma^2({D\over2}-1)\over(4\pi)^2\Gamma(D-2)}
\Big({4\pi\mu^2\over-\Lambda^2}\Big)^{2-D/2}
\;,\nonumber\\
&&B_{_0}(0,\;\Lambda^2,\;0)=B_{_0}(0,\;0,\;\Lambda^2)=-{i\Gamma(1-{D\over2})\over(4\pi)^2}
\Big({4\pi\mu^2\over\Lambda^2}\Big)^{2-D/2}
\;,\nonumber\\
&&B_{_0}(\Lambda^2,\;\Lambda^2,\;0)=B_{_0}(\Lambda^2,\;0,\;\Lambda^2)={i\Gamma(2-{D\over2})\over(4\pi)^2(D-3)}
\Big({4\pi\mu^2\over\Lambda^2}\Big)^{2-D/2}
\;,\nonumber\\
&&B_{_0}(0,\;\Lambda^2,\;\Lambda^2)={i\Gamma(2-{D\over2})\over(4\pi)^2}
\Big({4\pi\mu^2\over\Lambda^2}\Big)^{2-D/2}\;.
\label{GKZ33i}
\end{eqnarray}
Using above expressions, one derives the combination coefficients as
\begin{eqnarray}
&&A_{_{a,B}}=0,\;\;\;B_{_{a,B}}=C_{_{a,B}}=(-)^{D/2-1}\Gamma(1-{D\over2})
\;,\nonumber\\
&&D_{_{a,B}}=\Gamma^2({D\over2}-1)\Gamma(2-{D\over2})
\;,\nonumber\\
&&A_{_{b,B}}=A_{_{c,B}}=C_{_{b,B}}=C_{_{c,B}}=0
\;,\nonumber\\
&&B_{_{b,B}}=B_{_{c,B}}=-D_{_{b,B}}=-D_{_{c,B}}=\Gamma(1-{D\over2})\;.
\label{GKZ33j}
\end{eqnarray}
In a similar way, the combination coefficients involved in the Feynman integral of the two-loop vacuum
diagram are written as
\begin{eqnarray}
&&A_{_{a,V}}=-B_{_{a,V}}=-C_{_{a,V}}=\Gamma(3-{D\over2})\Gamma^2(1-{D\over2})
\;,\nonumber\\
&&D_{_{a,V}}=-2\Gamma^2(3-{D\over2})\Gamma({D\over2}-1)\Gamma(2-D)
\;,\nonumber\\
&&A_{_{b,V}}=-B_{_{b,V}}=D_{_{b,V}}=A_{_{a,V}},\;C_{_{b,V}}=D_{_{a,V}}
\;,\nonumber\\
&&A_{_{c,V}}=-B_{_{c,V}}=D_{_{c,V}}=A_{_{a,V}},\;C_{_{c,V}}=D_{_{a,V}}\;.
\label{GKZ33k}
\end{eqnarray}

For the triangulation $\triangle_{_\omega}=\{\sigma_{_1}^a,\;\sigma_{_2}^a,\;
\sigma_{_3}^a,\;\sigma_{_4}^a\}$ of the massless one-loop triangle diagram,
the canonical series solutions are constructed as
\begin{eqnarray}
&&\phi_{p_{_{\sigma_{_1}^a}}}=z_{_{C,1}}^{D/2-2}z_{_{C,2}}^{D/2-2}z_{_{C,3}}^{3-D}z_{_{C,4}}^{1-D/2}F_{_4}
\left(\left.\begin{array}{cc}D-3,&{D\over2}-1\\
{D\over2}-1,&{D\over2}-1\end{array}\right|{z_{_{C,1}}z_{_{C,5}}\over
z_{_{C,3}}z_{_{C,4}}},\;{z_{_{C,2}}z_{_{C,6}}\over
z_{_{C,3}}z_{_{C,4}}}\right)
\;,\nonumber\\
&&\phi_{p_{_{\sigma_{_2}^a}}}=z_{_{C,1}}^{D/2-2}z_{_{C,3}}^{1-D/2}z_{_{C,4}}^{-1}z_{_{C,6}}^{2-D/2}F_{_4}
\left(\left.\begin{array}{cc}1,&{D\over2}-1\\
{D\over2}-1,&3-{D\over2}\end{array}\right|{z_{_{C,1}}z_{_{C,5}}\over
z_{_{C,3}}z_{_{C,4}}},\;{z_{_{C,2}}z_{_{C,6}}\over
z_{_{C,3}}z_{_{C,4}}}\right)
\;,\nonumber\\
&&\phi_{p_{_{\sigma_{_3}^a}}}=z_{_{C,2}}^{D/2-2}z_{_{C,3}}^{1-D/2}z_{_{C,4}}^{-1}z_{_{C,5}}^{2-D/2}F_{_4}
\left(\left.\begin{array}{cc}1,&{D\over2}-1\\
3-{D\over2},&{D\over2}-1\end{array}\right|{z_{_{C,1}}z_{_{C,5}}\over
z_{_{C,3}}z_{_{C,4}}},\;{z_{_{C,2}}z_{_{C,6}}\over
z_{_{C,3}}z_{_{C,4}}}\right)
\;,\nonumber\\
&&\phi_{p_{_{\sigma_{_4}^a}}}=z_{_{C,3}}^{-1}z_{_{C,4}}^{D/2-3}z_{_{C,5}}^{2-D/2}z_{_{C,6}}^{2-D/2}F_{_4}
\left(\left.\begin{array}{cc}1,&3-{D\over2}\\
3-{D\over2},&3-{D\over2}\end{array}\right|{z_{_{C,1}}z_{_{C,5}}\over
z_{_{C,3}}z_{_{C,4}}},\;{z_{_{C,2}}z_{_{C,6}}\over
z_{_{C,3}}z_{_{C,4}}}\right)\;.
\label{GKZ34}
\end{eqnarray}
Similarly the canonical series solutions corresponding to
the triangulation $\triangle_{_\omega}=\{\sigma_{_1}^b,\;\sigma_{_2}^b,\;
\sigma_{_3}^b,\;\sigma_{_4}^b\}$ are written as
\begin{eqnarray}
&&\phi_{p_{_{\sigma_{_1}^b}}}=z_{_{C,1}}^{-1}z_{_{C,2}}^{D/2-2}z_{_{C,3}}^{2-D/2}z_{_{C,5}}^{1-D/2}F_{_4}
\left(\left.\begin{array}{cc}1,&{D\over2}-1\\
3-{D\over2},&{D\over2}-1\end{array}\right|{z_{_{C,3}}z_{_{C,4}}\over
z_{_{C,1}}z_{_{C,5}}},\;{z_{_{C,2}}z_{_{C,6}}\over
z_{_{C,1}}z_{_{C,5}}}\right)
\;,\nonumber\\
&&\phi_{p_{_{\sigma_{_2}^b}}}=z_{_{C,1}}^{1-D/2}z_{_{C,2}}^{D/2-2}z_{_{C,3}}^{D/2-2}z_{_{C,4}}^{3-D}F_{_4}
\left(\left.\begin{array}{cc}D-3,&{D\over2}-1\\
{D\over2}-1,&{D\over2}-1\end{array}\right|{z_{_{C,3}}z_{_{C,4}}\over
z_{_{C,1}}z_{_{C,5}}},\;{z_{_{C,2}}z_{_{C,6}}\over
z_{_{C,1}}z_{_{C,5}}}\right)
\;,\nonumber\\
&&\phi_{p_{_{\sigma_{_3}^b}}}=z_{_{C,1}}^{D/2-3}z_{_{C,3}}^{2-D/2}z_{_{C,5}}^{-1}z_{_{C,6}}^{2-D/2}F_{_4}
\left(\left.\begin{array}{cc}1,&3-{D\over2}\\
3-{D\over2},&3-{D\over2}\end{array}\right|{z_{_{C,3}}z_{_{C,4}}\over
z_{_{C,1}}z_{_{C,5}}},\;{z_{_{C,2}}z_{_{C,6}}\over
z_{_{C,1}}z_{_{C,5}}}\right)
\;,\nonumber\\
&&\phi_{p_{_{\sigma_{_4}^b}}}=z_{_{C,1}}^{-1}z_{_{C,4}}^{D/2-2}z_{_{C,5}}^{1-D/2}z_{_{C,6}}^{2-D/2}F_{_4}
\left(\left.\begin{array}{cc}1,&{D\over2}-1\\
{D\over2}-1,&3-{D\over2}\end{array}\right|{z_{_{C,3}}z_{_{C,4}}\over
z_{_{C,1}}z_{_{C,5}}},\;{z_{_{C,2}}z_{_{C,6}}\over
z_{_{C,1}}z_{_{C,5}}}\right)\;.
\label{GKZ35}
\end{eqnarray}
For the triangulation $\triangle_{_\omega}=\{\sigma_{_1}^c,\;\sigma_{_2}^c,\;
\sigma_{_3}^c,\;\sigma_{_4}^c\}$, the canonical series solutions are
\begin{eqnarray}
&&\phi_{p_{_{\sigma_{_1}^c}}}=z_{_{C,1}}^{D/2-2}z_{_{C,2}}^{-1}z_{_{C,3}}^{2-D/2}z_{_{C,6}}^{1-D/2}F_{_4}
\left(\left.\begin{array}{cc}1,&{D\over2}-1\\
{D\over2}-1,&3-{D\over2}\end{array}\right|{z_{_{C,1}}z_{_{C,5}}\over
z_{_{C,2}}z_{_{C,6}}},\;{z_{_{C,3}}z_{_{C,4}}\over
z_{_{C,2}}z_{_{C,6}}}\right)
\;,\nonumber\\
&&\phi_{p_{_{\sigma_{_2}^c}}}=z_{_{C,1}}^{D/2-2}z_{_{C,2}}^{1-D/2}z_{_{C,4}}^{D/2-2}z_{_{C,6}}^{3-D}F_{_4}
\left(\left.\begin{array}{cc}D-3,&{D\over2}-1\\
{D\over2}-1,&{D\over2}-1\end{array}\right|{z_{_{C,1}}z_{_{C,5}}\over
z_{_{C,2}}z_{_{C,6}}},\;{z_{_{C,3}}z_{_{C,4}}\over
z_{_{C,2}}z_{_{C,6}}}\right)
\;,\nonumber\\
&&\phi_{p_{_{\sigma_{_3}^c}}}=z_{_{C,2}}^{D/2-3}z_{_{C,3}}^{2-D/2}z_{_{C,5}}^{2-D/2}z_{_{C,6}}^{-1}F_{_4}
\left(\left.\begin{array}{cc}1,&3-{D\over2}\\
3-{D\over2},&3-{D\over2}\end{array}\right|{z_{_{C,1}}z_{_{C,5}}\over
z_{_{C,2}}z_{_{C,6}}},\;{z_{_{C,3}}z_{_{C,4}}\over
z_{_{C,2}}z_{_{C,6}}}\right)
\;,\nonumber\\
&&\phi_{p_{_{\sigma_{_4}^c}}}=z_{_{C,2}}^{-1}z_{_{C,4}}^{D/2-2}z_{_{C,5}}^{2-D/2}z_{_{C,6}}^{1-D/2}F_{_4}
\left(\left.\begin{array}{cc}1,&{D\over2}-1\\
3-{D\over2},&{D\over2}-1\end{array}\right|{z_{_{C,1}}z_{_{C,5}}\over
z_{_{C,2}}z_{_{C,6}}},\;{z_{_{C,3}}z_{_{C,4}}\over
z_{_{C,2}}z_{_{C,6}}}\right)\;.
\label{GKZ36}
\end{eqnarray}

Similarly setting $z_{_{C,1}}=x_{_C}$, $z_{_{C,2}}=y_{_C}$, and $z_{_{C,k}}=1$ with
$k=3,\cdots,6$, one formulates the Feynman integral as the linear combinations
of canonical series solutions in the corresponding parameter space.
\begin{itemize}
\item For $|x_{_C}|\le1,\;|y_{_C}|\le1$, the Feynman integral is
\begin{eqnarray}
&&S_{_{a,C}}(x_{_C},y_{_C})=A_{_{a,C}}(x_{_{C}}y_{_{C}})^{D/2-2}F_{_4}
\left(\left.\begin{array}{cc}D-3,&{D\over2}-1\\
{D\over2}-1,&{D\over2}-1\end{array}\right|x_{_{C}},\;y_{_{C}}\right)
\nonumber\\
&&\hspace{2.2cm}
+B_{_{a,C}}x_{_{i}}^{D/2-1}F_{_4}
\left(\left.\begin{array}{cc}1,&{D\over2}-1\\
{D\over2}-1,&3-{D\over2}\end{array}\right|x_{_{C}},\;y_{_{C}}\right)
\nonumber\\
&&\hspace{2.2cm}
+C_{_{a,C}}y_{_{i}}^{D/2-1}F_{_4}\left(\left.\begin{array}{cc}1,&{D\over2}-1\\
3-{D\over2},&{D\over2}-1\end{array}\right|x_{_{C}},\;y_{_{C}}\right)
\nonumber\\
&&\hspace{2.2cm}
+D_{_{a,C}}F_{_4}\left(\left.\begin{array}{cc}1,&3-{D\over2}\\
3-{D\over2},&3-{D\over2}\end{array}\right|x_{_{C}},\;y_{_{C}}\right)\;,
\label{GKZ36a}
\end{eqnarray}
which is convergent in the region $\sqrt{|x_{_{C}}|}+\sqrt{|y_{_{C}}|}<1$.

\item For $|x_{_C}|\ge1,\;|y_{_C}|\le1$, the Feynman integral is
\begin{eqnarray}
&&S_{_{b,C}}(x_{_C},y_{_C})=A_{_{b,C}}x_{_{C}}^{-1}y_{_{C}}^{D/2-2}F_{_4}
\left(\left.\begin{array}{cc}1,&{D\over2}-1\\
3-{D\over2},&{D\over2}-1\end{array}\right|{1\over x_{_{C}}},\;{y_{_{C}}\over x_{_C}}\right)
\nonumber\\
&&\hspace{2.2cm}
+B_{_{b,C}}x_{_{C}}^{1-D/2}y_{_C}^{D/2-2}F_{_4}\left(\left.\begin{array}{cc}D-3,&{D\over2}-1\\
{D\over2}-1,&{D\over2}-1\end{array}\right|{1\over x_{_{C}}},\;{y_{_{C}}\over x_{_C}}\right)
\nonumber\\
&&\hspace{2.2cm}
+C_{_{b,C}}x_{_{C}}^{D/2-3}F_{_4}\left(\left.\begin{array}{cc}1,&3-{D\over2}\\
3-{D\over2},&3-{D\over2}\end{array}\right|{1\over x_{_{C}}},\;{y_{_{C}}\over x_{_C}}\right)
\nonumber\\
&&\hspace{2.2cm}
+D_{_{b,C}}x_{_{C}}^{-1}F_{_4}\left(\left.\begin{array}{cc}1,&{D\over2}-1\\
{D\over2}-1,&3-{D\over2}\end{array}\right|{1\over x_{_{C}}},\;{y_{_{C}}\over x_{_C}}\right)
\label{GKZ36b}
\end{eqnarray}
which is convergent in the region $1+\sqrt{|y_{_{C}}|}<\sqrt{|x_{_{C}}|}$.

\item For $|x_{_C}|\le1,\;|y_{_C}|\ge1$, the Feynman integral is
\begin{eqnarray}
&&S_{_{c,C}}(x_{_C},y_{_C})=A_{_{c,C}}x_{_{C}}^{D/2-2}y_{_{C}}^{-1}F_{_4}
\left(\left.\begin{array}{cc}1,&{D\over2}-1\\{D\over2}-1,&3-{D\over2}\end{array}\right|
{x_{_{C}}\over y_{_{C}}},\;{1\over y_{_{C}}}\right)
\nonumber\\
&&\hspace{2.2cm}
+B_{_{c,C}}x_{_{C}}^{D/2-2}y_{_{C}}^{1-D/2}F_{_4}
\left(\left.\begin{array}{cc}D-3,&{D\over2}-1\\{D\over2}-1,&{D\over2}-1\end{array}\right|
{x_{_{C}}\over y_{_{C}}},\;{1\over y_{_{C}}}\right)
\nonumber\\
&&\hspace{2.2cm}
+C_{_{c,C}}y_{_{C}}^{D/2-3}F_{_4}
\left(\left.\begin{array}{cc}1,&3-{D\over2}\\3-{D\over2},&3-{D\over2}\end{array}\right|
{x_{_{C}}\over y_{_{C}}},\;{1\over y_{_{C}}}\right)
\nonumber\\
&&\hspace{2.2cm}
+D_{_{c,C}}y_{_{C}}^{-1}F_{_4}
\left(\left.\begin{array}{cc}1,&{D\over2}-1\\3-{D\over2},&{D\over2}-1\end{array}\right|
{x_{_{C}}\over y_{_{C}}},\;{1\over y_{_{C}}}\right)\;,
\label{GKZ36c}
\end{eqnarray}
which is convergent in the region $1+\sqrt{|x_{_{C}}|}<\sqrt{|y_{_{C}}|}$.
\end{itemize}
Using the expressions of Feynman integral of the massless triangle diagram at some special
kinematic points $(\Lambda^2,\;\Lambda^2,\;0),\;(\Lambda^2,\;0,\;0)$ etc, one obtains those integration
constants as
\begin{eqnarray}
&&A_{_{a,C}}=-\Gamma^2(2-{D\over2})\Gamma({D\over2}-1)
\;,\nonumber\\
&&B_{_{a,C}}=C_{_{a,C}}=-D_{_{a,C}}={\Gamma(2-{D\over2})\Gamma^2({D\over2}-1)\over
(2-{D\over2})\Gamma(D-3)}
\;,\nonumber\\
&&B_{_{b,C}}=B_{_{c,C}}=A_{_{a,C}}\;,\;\;D_{_{b,C}}=A_{_{c,C}}=B_{_{a,C}}
\;,\nonumber\\
&&A_{_{b,C}}=D_{_{c,C}}=C_{_{a,C}},\;\;C_{_{b,C}}=C_{_{c,C}}=D_{_{a,C}}\;.
\label{GKZ36d}
\end{eqnarray}

Actually Feynman integrals presented here can be written in terms of Gauss
function~\cite{Davydychev2000} by the well-known reduction of
Appell function of the fourth kind~\cite{Davydychev1993NPB},
then the analytic continuation of those Feynman integrals is made to the whole parameter space
through the transformations of Gauss functions.

\section{GKZ-hypergeometric systems of other Feynman integrals\label{sec4}}
\subsection{Sunset diagram with three differential masses\label{sec4-1}}
\indent\indent
In order to make the notation less cluttered, we adopt the multi-index convention~\cite{Coutinho1995},
and write Feynman integral of the two-loop sunset diagram as
\begin{eqnarray}
&&\Sigma_{_{\ominus}}(p^2)=
-{p^2\Gamma^2(3-{D\over2})\over(4\pi)^4}\Big({4\pi\mu^2\over-p^2}\Big)^{4-D}
T_{_{123}}^{p}({\bf a},\;{\bf b};\;{\bf x})\;,
\label{SunSet-GKZ1}
\end{eqnarray}
with the multi-index notations ${\bf a}=(a_{_1},a_{_2})$,
${\bf b}=(b_{_1},b_{_2},b_{_3})$, and ${\bf x}=(x_{_1},x_{_2},x_{_3})$.
Where $a_{_1}=3-D$, $a_{_2}=4-3D/2$, $b_{_1}=b_{_2}=b_{_3}=2-D/2$,
and $x_{_1}=m_{_1}^2/p^2$, $x_{_2}=m_{_2}^2/p^2$, $x_{_3}=m_{_3}^2/p^2$.
Certainly the dimensionless function $T_{_{123}}^{p}$
complies with the third Lauricella's system of linear PDEs~\cite{Feng2018,Feng2019}
\begin{eqnarray}
&&\Big\{\hat{\vartheta}_{x_{_k}}(\hat{\vartheta}_{x_{_k}}+b_{_k}-1)
-x_{_k}(\sum\limits_{i=1}^3\hat{\vartheta}_{x_{_i}}+a_{_1})
(\sum\limits_{i=1}^3\hat{\vartheta}_{x_{_i}}+a_{_2})\Big\}T_{_{123}}^{p}=0\;,
(k=1,\;2,\;3)\;.
\label{SunSet-GKZ2}
\end{eqnarray}
Using the system of linear PDEs,
one derives the following relations between $T_{_{123}}^{p}$ and its contiguous
functions as
\begin{eqnarray}
&&(\sum\limits_{i=1}^3\hat{\vartheta}_{x_{_i}}+a_{_j})
T_{_{123}}^{p}({\bf a},\;{\bf b};\;\;{\bf x})=
a_{_j}T_{_{123}}^{p}({\bf a}+{\bf n}_{_{2,j}},\;{\bf b};\;\;{\bf x}),\;(j=1,\;2)
\;,\nonumber\\
&&(\hat{\vartheta}_{_{x_{_k}}}+b_{_k}-1)
T_{_{123}}^{p}({\bf a},\;{\bf b};\;\;{\bf x})=
(b_{_k}-1)T_{_{123}}^{p}({\bf a},\;{\bf b}-{\bf n}_{_{3,k}};\;\;{\bf x})
\;,\nonumber\\
&&\partial_{_{x_{_k}}}T_{_{123}}^{p}({\bf a},\;{\bf b};\;\;{\bf x})=
{a_{_1}a_{_2}\over b_{_k}}T_{_{123}}^{p}({\bf a}+{\bf n}_{_2},\;
{\bf b}+{\bf n}_{_{3,k}};\;\;{\bf x})\;,(k=1,\;2,\;3)\;.
\label{SunSet-GKZ3}
\end{eqnarray}
Where ${\bf n}_{_{2,j}}\in{\bf R}^2,\;(j=1,2)$ denotes the row vector whose entry is zero
except that the $j-$th entry is $1$, and the row vector ${\bf n}_{_2}=(1,1)$.
In addition ${\bf n}_{_{3,k}}\in{\bf R}^3,\;(k=1,2,3)$ denotes the row vector whose entries are zero
except that the $j-$th entry is $1$ and ${\bf n}_{_3}=(1,1,1)$.
Following the work of W.~Miller~\cite{Miller68,Miller72}, we define the auxiliary function $\Phi$ as
\begin{eqnarray}
&&\Phi({\bf a},\;{\bf b};\;{\bf x},\;{\bf u},\;{\bf v})=
{\bf u}^{\bf a}\;{\bf v}^{{\bf b}-{\bf n}_{_3}}\;T_{_{123}}^{p}({\bf a},\;{\bf b};\;\;{\bf x})\;.
\label{SunSet-GKZ4}
\end{eqnarray}
Where the row vectors ${\bf u}=(u_{_1},u_{_2})\in{\bf R}^2$, ${\bf v}=(v_{_1},v_{_2},v_{_3})\in{\bf R}^3$,
and the multi-index notations ${\bf u}^{\bf a}=u_{_1}^{a_{_1}}u_{_2}^{a_{_2}}$,
${\bf v}^{{\bf b}-{\bf n}_{_3}}=\prod\limits_{i=1}^3v_{_i}^{b_{_i}-1}$.
Miller's transformation~\cite{Miller68,Miller72} on the function $T_{_{123}}^{p}$ is to replace
the multiplication by the parameter $a_{_j},\;b_{_k}$ in Eq.~(\ref{SunSet-GKZ3}) by Euler operators
$\hat{\vartheta}_{_{u_{_j}}}$, $\hat{\vartheta}_{_{v_{_k}}}$:
\begin{eqnarray}
&&\hat{\vartheta}_{_{u_{_j}}}\Phi({\bf a},\;{\bf b};\;{\bf x},\;{\bf u},\;{\bf v})=
a_{_j}\Phi({\bf a},\;{\bf b};\;{\bf x},\;{\bf u},\;{\bf v}),\;(j=1,2)
\;,\nonumber\\
&&\hat{\vartheta}_{_{v_{_k}}}\Phi({\bf a},\;{\bf b};\;{\bf x},\;{\bf u},\;{\bf v})=
(b_{_k}-1)\Phi({\bf a},\;{\bf b};\;{\bf x},\;{\bf u},\;{\bf v}),\;(k=1,2,3)\;,
\label{SunSet-GKZ5}
\end{eqnarray}
which induces the notion of GKZ-hypergeometric system naturally.
In addition, the contiguous relations of the function defined in Eq.~(\ref{SunSet-GKZ4}) are given as
\begin{eqnarray}
&&\hat{\cal O}_{_1}\Phi({\bf a},\;{\bf b};\;{\bf x},\;{\bf u},\;{\bf v})=
a_{_j}\Phi({\bf a}+{\bf n}_{_{2,j}},\;{\bf b};\;{\bf x},\;{\bf u},\;{\bf v}),\;(j=1,2)
\;,\nonumber\\
&&\hat{\cal O}_{_{2+k}}\Phi({\bf a},\;{\bf b};\;{\bf x},\;{\bf u},\;{\bf v})=
(b_{_k}-1)\Phi({\bf a},\;{\bf b}-{\bf n}_{_{3,k}};\;{\bf x},\;{\bf u},\;{\bf v})
\;,\nonumber\\
&&\hat{\cal O}_{_{5+k}}\Phi({\bf a},\;{\bf b};\;{\bf x},\;{\bf u},\;{\bf v})=
{a_{_1}a_{_2}\over b_{_k}}\Phi({\bf a}+{\bf n}_{_2},\;{\bf b}+{\bf n}_{_{3,k}};\;{\bf x},\;{\bf u},\;{\bf v})
,\;(k=1,2,3)\;,
\label{SunSet-GKZ6}
\end{eqnarray}
where the operators $\hat{\cal O}_{_n}^i(n=1,\cdots,8)$ are
\begin{eqnarray}
&&\hat{\cal O}_{_1}=u_{_1}(\sum\limits_{i=1}^3\hat{\vartheta}_{_{x_{_i}}}+\hat{\vartheta}_{_{u_{_1}}})
\;,\nonumber\\
&&\hat{\cal O}_{_2}=u_{_2}(\sum\limits_{i=1}^3\hat{\vartheta}_{_{x_{_i}}}+\hat{\vartheta}_{_{u_{_2}}})
\;,\nonumber\\
&&\hat{\cal O}_{_{2+k}}={1\over v_{_k}}(\hat{\vartheta}_{_{x_{_k}}}+\hat{\vartheta}_{_{v_{_k}}})
\;,\nonumber\\
&&\hat{\cal O}_{_{5+k}}=u_{_1}u_{_2}v_{_k}\partial_{_{x_{_k}}}
\;,(k=1,\;2,\;3)\;.
\label{SunSet-GKZ7}
\end{eqnarray}
Those operators together with $\hat{\vartheta}_{_{u_{_j}}},\;\hat{\vartheta}_{_{v_{_k}}}$ define the Lie algebra of the
hypergeometric system~\cite{Miller68,Miller72} in Eq.(\ref{SunSet-GKZ2}). Through the transformation of indeterminates
\begin{eqnarray}
&&z_{_1}={1\over u_{_1}}\;,\;\;z_{_2}={1\over u_{_2}}\;,
\nonumber\\
&&z_{_{2+k}}=v_{_i}\;,z_{_{5+k}}={x_{_k}\over u_{_1}u_{_2}v_{_k}}\;,\;\;(k=1,\;2,\;3)\;,
\label{SunSet-GKZ8}
\end{eqnarray}
the equations in Eq.~(\ref{SunSet-GKZ5}) are changed as
\begin{eqnarray}
&&\Big(\mathbf{A}_{_{\ominus}}\cdot\vec{\vartheta}\Big)\Phi({\bf a},\;{\bf b};\;{\bf x},\;{\bf u},\;{\bf v})
=\mathbf{B}_{_{\ominus}}\Phi({\bf a},\;{\bf b};\;{\bf x},\;{\bf u},\;{\bf v})\;,
\label{SunSet-GKZ9}
\end{eqnarray}
where
\begin{eqnarray}
&&\mathbf{A}_{_{\ominus}}=\left(\begin{array}{cccccccc}1\;&0\;&0\;&0\;&0\;&1\;&1\;&1\;\\0\;&1\;&0\;&0\;&0\;&1\;&1\;&1\;\\
0\;&0\;&1\;&0\;&0\;&-1\;&0\;&0\;\\0\;&0\;&0\;&1\;&0\;&0\;&-1\;&0\;\\0\;&0\;&0\;&0\;&1\;&0\;&0\;&-1\;
\end{array}\right)
\;,\nonumber\\
&&\vec{\vartheta}^{\;T}=(\vartheta_{_{z_{_1}}},\;\vartheta_{_{z_{_2}}},\;\vartheta_{_{z_{_3}}},\;
\vartheta_{_{z_{_4}}},\;\vartheta_{_{z_{_5}}},\;\vartheta_{_{z_{_6}}},\;\vartheta_{_{z_{_7}}},\;\vartheta_{_{z_{_8}}})
\;,\nonumber\\
&&\mathbf{B}_{_{\ominus}}^{\;T}=(-a_{_1},\;-a_{_2},\;b_{_1}-1,\;b_{_2}-1,\;b_{_3}-1)\;.
\label{SunSet-GKZ10}
\end{eqnarray}
Correspondingly the universal Gr\"obner basis of the toric ideal associated with $\mathbf{A}_{_{\ominus}}$ is
\begin{eqnarray}
&&{\cal U}_{_{\mathbf{A}_{_{\ominus}}}}=\{\partial_{_{z_{_1}}}\partial_{_{z_{_2}}}-\partial_{_{z_{_3}}}\partial_{_{z_{_6}}},\;
\partial_{_{z_{_1}}}\partial_{_{z_{_2}}}-\partial_{_{z_{_4}}}\partial_{_{z_{_7}}},\;
\partial_{_{z_{_1}}}\partial_{_{z_{_2}}}-\partial_{_{z_{_5}}}\partial_{_{z_{_8}}},\;
\nonumber\\
&&\hspace{1.2cm}
\partial_{_{z_{_3}}}\partial_{_{z_{_6}}}-\partial_{_{z_{_4}}}\partial_{_{z_{_7}}},\;
\partial_{_{z_{_3}}}\partial_{_{z_{_6}}}-\partial_{_{z_{_5}}}\partial_{_{z_{_8}}},\;
\partial_{_{z_{_4}}}\partial_{_{z_{_7}}}-\partial_{_{z_{_5}}}\partial_{_{z_{_8}}}\}\;.
\label{SunSet-GKZ11}
\end{eqnarray}
The operators $\mathbf{A}_{_{\ominus}}\cdot\vec{\vartheta}-\mathbf{B}_{_{\ominus}}$ and that from the set
${\cal U}_{_{\mathbf{A}_{_{\ominus}}}}$ compose the generators of
a left ideal in the Weyl algebra $D={\bf C}\langle z_{_{1}},\;\cdots,\;z_{_{8}},\;\partial_{_{z_{_{1}}}},\;
\cdots,\;\partial_{_{z_{_{8}}}}\rangle$. Defining an isomorphism between the commutative
polynomial ring and the Weyl algebra~\cite{M.Saito2000}
\begin{eqnarray}
&&\Psi:\;\;{\bf C}[z_{_{1}},\;\cdots,\;z_{_{8}},\;\xi_{_{1}},\;\cdots,\;\xi_{_{8}}]
\rightarrow\;D,\;z_{_i}^\alpha\xi_{_i}^\beta\mapsto z_{_i}^\alpha\partial_{_{z_{_{i}}}}^\beta\;,
\label{SunSet-GKZ11a}
\end{eqnarray}
one obtains the state polytope~\cite{Sturmfels1995} of the preimage of the universal Gr\"obner basis
in Eq.~(\ref{SunSet-GKZ11})
\begin{eqnarray}
&&\xi_{_{6}}+\xi_{_{7}}+\xi_{_{8}}\ge3,\;\xi_{_{6}}+\xi_{_{7}}\ge1,\;
\xi_{_{6}}+\xi_{_{8}}\ge1,
\nonumber\\
&&\xi_{_{7}}+\xi_{_{8}}\ge1,\;\xi_{_{6}}\ge0,\;\xi_{_{7}}\ge0,\;\xi_{_{8}}\ge0,\;-\xi_{_{8}}\ge-3,
\nonumber\\
&&-\xi_{_{7}}\ge-3,\;-\xi_{_{6}}\ge-3,\;-\xi_{_{7}}-\xi_{_{8}}\ge-5,\;-\xi_{_{6}}-\xi_{_{8}}\ge-5,
\nonumber\\
&&-\xi_{_{6}}-\xi_{_{7}}\ge-5,,\;-\xi_{_{6}}-\xi_{_{7}}-\xi_{_{8}}\ge-6\;,
\label{SunSet-GKZ11b}
\end{eqnarray}
on the hyperplane
\begin{eqnarray}
&&\xi_{_{1}}=\xi_{_{2}},\;\;\;\xi_{_{3}}=\xi_{_{6}},\;\;\;\xi_{_{4}}=\xi_{_{7}},\nonumber\\
&&\xi_{_{5}}=\xi_{_{8}},\;\;\;\xi_{_{2}}+\xi_{_{6}}+\xi_{_{7}}+\xi_{_{8}}=6\;.
\label{SunSet-GKZ11c}
\end{eqnarray}
The normal fan of the state polytope in Eq.~(\ref{SunSet-GKZ11b}) is the Gr\"obner fan
of the left ideal generated by the operators in Eq.~(\ref{SunSet-GKZ9}) and Eq.~(\ref{SunSet-GKZ11}).
Because codimension$=3$ for GKZ-hypergeometric system here, the Gr\"obner fan
refines the secondary fan which is composed by the column vectors of a Gale transform
of the matrix $A_{_{\ominus}}$ in Eq.(\ref{SunSet-GKZ10}). With these fans, one constructs canonical basis series solutions
in C-type Lauricella functions with three variables~\cite{Feng2019,Berends1994,Ananthanarayan2019}.
In order to make analytic continuation of Lauricella functions from their convergent regions
to the whole parameter space, we should perform some linear fractional transformations
among the complex variables $z_{_{1}},\cdots,z_{_{8}}$. We will release our calculation results
further elsewhere.

\subsection{$C_{_0}$ function with one nonzero mass\label{sec4-2}}
In this case, the scalar integral
\begin{eqnarray}
&&C_{_0}^a(p_{_1}^2,\;p_{_2}^2,\;p_{_3}^2,\;m^2)=
\int{d^Dq\over(2\pi)^D}{1\over(q^2-m^2)(q+p_{_1})^2(q-p_{_2})^2}
\nonumber\\
&&\hspace{3.5cm}=
{i(-)^{D/2}(p_{_3}^2)^{D/2-3}\over(4\pi)^{D/2}}
F_{_{a,p_{_3}}}({\bf a},\;{\bf b};\;\;{\bf x})
\label{GKZ-1M-1}
\end{eqnarray}
where the row vectors ${\bf a}=(a_{_1},a_{_2},a_{_3})\in {\bf R}^3$, ${\bf b}=(b_{_1},b_{_2})\in{\bf R}^2$,
and ${\bf x}=(\xi_{_{33}},x_{_{13}},x_{_{23}})\in{\bf R}^3$, respectively. Additionally the parameters $a_{_1}=4-D$,
$a_{_2}=3-D/2$, $a_{_3}=1$, $b_{_1}=b_{_2}=3-D/2$, $p_{_3}^2=(p_{_1}+p_{_2})^2$,
and the dimensionless ratios $\xi_{_{33}}=-m^2/p_{_3}^2$, $x_{_{ij}}=p_{_i}^2/p_{_j}^2,\;(i,\;j=1,2,3)$.
The dimensionless function $F_{_{a,p_{_3}}}$ satisfies the holonomic hypergeometric system
of linear PDEs
\begin{eqnarray}
&&\Big\{(a_{_1}+\hat{\vartheta}_{\xi_{_{33}}})
(a_{_2}+\hat{\vartheta}_{\xi_{_{33}}}+\sum\limits_{i=1}^2\hat{\vartheta}_{x_{_{i3}}})
(a_{_3}+\hat{\vartheta}_{\xi_{_{33}}}+\sum\limits_{i=1}^2\hat{\vartheta}_{x_{_{i3}}})
\nonumber\\
&&\hspace{0.0cm}
-{1\over\xi_{_{33}}}\hat{\vartheta}_{\xi_{_{33}}}\prod\limits_{i=1}^2
(b_{_i}+\hat{\vartheta}_{\xi_{_{33}}}+\hat{\vartheta}_{x_{_{i3}}})\Big\}F_{_{a,p_{_3}}}=0
\;,\nonumber\\
&&\Big\{(a_{_2}+\hat{\vartheta}_{\xi_{_{33}}}+\sum\limits_{i=1}^2\hat{\vartheta}_{x_{_{i3}}})
(a_{_3}+\hat{\vartheta}_{\xi_{_{33}}}+\sum\limits_{i=1}^2\hat{\vartheta}_{x_{_{i3}}})
\nonumber\\
&&\hspace{0.0cm}
-{1\over x_{_{j3}}}\hat{\vartheta}_{x_{_{j3}}}(b_{_j}+\hat{\vartheta}_{\xi_{_{33}}}
+\hat{\vartheta}_{x_{_{j3}}})\Big\}F_{_{a,p_{_3}}}=0\;,\;\;(j=1,\;2)\;.
\label{GKZ-1M-2}
\end{eqnarray}
Defining Miller's transformation on the function $F_{_{a,p_{_3}}}$
\begin{eqnarray}
&&\Phi_{_{a,p_{_3}}}({\bf a},\;{\bf b};\;\;{\bf x},\;{\bf u},\;{\bf v})
={\bf u}^{\bf a}{\bf v}^{{\bf b}-{\bf n}_{_2}}
F_{_{a,p_{_3}}}({\bf a},\;{\bf b};\;\;{\bf x})
\label{GKZ-1M-3}
\end{eqnarray}
with ${\bf u}=(u_{_1},u_{_2},u_{_3})\in{\bf R}^3$, ${\bf v}=(v_{_1},v_{_2})\in{\bf R}^2$,
one replaces the multiplication of the parameter $a_{_k},\;b_{_j}$ by Euler operators
$\hat{\vartheta}_{_{u_{_k}}}$, $\hat{\vartheta}_{_{v_{_j}}}$:
\begin{eqnarray}
&&\hat{\vartheta}_{_{u_{_k}}}\Phi_{_{a,p_{_3}}}({\bf a},\;{\bf b};\;\;{\bf x},\;{\bf u},\;{\bf v})=
a_{_k}\Phi_{_{a,p_{_3}}}({\bf a},\;{\bf b};\;\;{\bf x},\;{\bf u},\;{\bf v})
\;,\nonumber\\
&&\hat{\vartheta}_{_{v_{_j}}}\Phi_{_{a,p_{_3}}}({\bf a},\;{\bf b};\;\;{\bf x},\;{\bf u},\;{\bf v})=
(b_{_j}-1)\Phi_{_{a,p_{_3}}}({\bf a},\;{\bf b};\;\;{\bf x},\;{\bf u},\;{\bf v})
\label{GKZ-1M-4}
\end{eqnarray}
where $k=1,\;2,\;3$ and $j=1,\;2$, respectively.
In addition, the contiguous relations of the auxiliary function are given as
\begin{eqnarray}
&&\hat{\cal O}_{_k}\Phi_{_{a,p_{_3}}}({\bf a},\;{\bf b};\;\;{\bf x},\;{\bf u},\;{\bf v})=
a_{_{k}}\Phi_{_{a,p_{_3}}}({\bf a}+{\bf n}_{_{3,k}},\;{\bf b};\;\;{\bf x},\;{\bf u},\;{\bf v})
\;,\nonumber\\
&&\hat{\cal O}_{_{3+j}}\Phi_{_{a,p_{_3}}}({\bf a},\;{\bf b};\;\;{\bf x},\;{\bf u},\;{\bf v})=
(b_{_j}-1)\Phi_{_{a,p_{_3}}}({\bf a},\;{\bf b}-{\bf n}_{_{2,j}};\;\;{\bf x},\;{\bf u},\;{\bf v})
\;,\nonumber\\
&&\hat{\cal O}_{_6}\Phi_{_{a,p_{_3}}}({\bf a},\;{\bf b};\;\;{\bf x},\;{\bf u},\;{\bf v})=
{a_{_1}a_{_2}a_{_3}\over b_{_1}b_{_2}}
\Phi_{_{a,p_{_3}}}({\bf a}+{\bf n}_{_3},\;{\bf b}+{\bf n}_{_2};\;\;{\bf x},\;{\bf u},\;{\bf v})
\;,\nonumber\\
&&\hat{\cal O}_{_{6+j}}\Phi_{_{a,p_{_3}}}({\bf a},\;{\bf b};\;\;{\bf x},\;{\bf u},\;{\bf v})=
{a_{_{2}}a_{_{3}}\over b_{_{j}}}\Phi_{_{a,p_{_3}}}({\bf a}+{\bf n}_{_{3,2}}
+{\bf n}_{_{3,3}},\;{\bf b}+{\bf n}_{_{2,j}};\;\;{\bf x},\;{\bf u},\;{\bf v})\;,
\label{GKZ-1M-5}
\end{eqnarray}
where the operators $\hat{\cal O}_{_n}^i(n=1,\cdots,8)$ are defined as
\begin{eqnarray}
&&\hat{\cal O}_{_1}=u_{_1}(\hat{\vartheta}_{_{\xi_{_{33}}}}+\hat{\vartheta}_{_{u_{_1}}})
\;,\nonumber\\
&&\hat{\cal O}_{_2}=u_{_2}(\hat{\vartheta}_{_{\xi_{_{33}}}}
+\sum\limits_{i=1}^2\hat{\vartheta}_{_{x_{_{i3}}}}+\hat{\vartheta}_{_{u_{_2}}})
\;,\nonumber\\
&&\hat{\cal O}_{_3}=u_{_3}(\hat{\vartheta}_{_{\xi_{_{33}}}}
+\sum\limits_{i=1}^2\hat{\vartheta}_{_{x_{_{i3}}}}+\hat{\vartheta}_{_{u_{_3}}})
\;,\nonumber\\
&&\hat{\cal O}_{_{3+j}}={1\over v_{_j}}(\hat{\vartheta}_{_{\xi_{_{33}}}}
+\hat{\vartheta}_{_{x_{_{j3}}}}+\hat{\vartheta}_{_{v_{_j}}})\;\;(j=1,\;2)
\;,\nonumber\\
&&\hat{\cal O}_{_{6}}=u_{_1}u_{_2}u_{_3}v_{_1}v_{_2}\partial_{_{\xi_{_{33}}}}
\;,\nonumber\\
&&\hat{\cal O}_{_{6+j}}=u_{_2}u_{_3}v_{_j}\partial_{_{x_{_{j3}}}}
\;,(j=1,\;2)\;.
\label{GKZ-1M-6}
\end{eqnarray}
Those operators above together with $\hat{\vartheta}_{_{u_{_k}}},\;\hat{\vartheta}_{_{v_{_j}}}$ define the Lie algebra of the
hypergeometric system~\cite{Miller68,Miller72} in Eq.(\ref{GKZ-1M-2}). Through the transformation of indeterminates
\begin{eqnarray}
&&z_{_k}^a={1\over u_{_k}}\;,\;\;z_{_{3+j}}^a=v_{_j}\;,
\nonumber\\
&&z_{_{6}}^a={\xi_{_{33}}\over u_{_1}u_{_2}u_{_3}v_{_1}v_{_2}}\;,
z_{_{6+j}}^a={x_{_{j3}}\over u_{_2}u_{_3}v_{_j}}\;,\;\;(k=1,\;2,\;3,\;\;j=1,\;2)\;,
\label{GKZ-1M-7}
\end{eqnarray}
the equations in Eq.~(\ref{GKZ-1M-4}) are rewritten as
\begin{eqnarray}
&&\Big(\mathbf{A}_{_a}\cdot\vec{\vartheta}_{_a}\Big)\Phi_{_{a,p_{_3}}}({\bf a},\;{\bf b};\;\;{\bf x},\;{\bf u},\;{\bf v})
=\mathbf{B_{_a}}\Phi_{_{a,p_{_3}}}({\bf a},\;{\bf b};\;\;{\bf x},\;{\bf u},\;{\bf v})\;,
\label{GKZ-1M-8}
\end{eqnarray}
where
\begin{eqnarray}
&&\mathbf{A}_{_a}=\left(\begin{array}{cccccccc}1\;&0\;&0\;&0\;&0\;&1\;&0\;&0\;\\0\;&1\;&0\;&0\;&0\;&1\;&1\;&1\;\\
0\;&0\;&1\;&0\;&0\;&1\;&1\;&1\;\\0\;&0\;&0\;&1\;&0\;&-1\;&-1\;&0\;\\0\;&0\;&0\;&0\;&1\;&-1\;&0\;&-1\;
\end{array}\right)
\;,\nonumber\\
&&\vec{\vartheta}_{_a}^{\;T}=(\vartheta_{_{z_{_1}^a}},\;\vartheta_{_{z_{_2}^a}},\;\vartheta_{_{z_{_3}^a}},\;
\vartheta_{_{z_{_4}^a}},\;\vartheta_{_{z_{_5}^a}},\;\vartheta_{_{z_{_6}^a}},\;\vartheta_{_{z_{_7}^a}},\;\vartheta_{_{z_{_8}^a}})
\;,\nonumber\\
&&\mathbf{B}_{_a}^{\;T}=(-a_{_1},\;-a_{_2},\;-a_{_3},\;b_{_1},\;b_{_2})\;.
\label{GKZ-1M-9}
\end{eqnarray}
Correspondingly the universal Gr\"obner basis of the toric ideal associated with $\mathbf{A}_a$ is
\begin{eqnarray}
&&{\cal U}_{_{\mathbf{A}_a}}=\{\partial_{_{z_{_1}^a}}\partial_{_{z_{_7}^a}}-\partial_{_{z_{_5}^a}}\partial_{_{z_{_6}^a}},\;
\partial_{_{z_{_1}^a}}\partial_{_{z_{_8}^a}}-\partial_{_{z_{_4}^a}}\partial_{_{z_{_6}^a}},\;
\partial_{_{z_{_2}^a}}\partial_{_{z_{_3}^a}}-\partial_{_{z_{_4}^a}}\partial_{_{z_{_7}^a}},\;
\partial_{_{z_{_2}^a}}\partial_{_{z_{_3}^a}}-\partial_{_{z_{_5}^a}}\partial_{_{z_{_8}^a}},\;
\nonumber\\
&&\hspace{1.2cm}
\partial_{_{z_{_4}^a}}\partial_{_{z_{_7}^a}}-\partial_{_{z_{_5}^a}}\partial_{_{z_{_8}^a}},\;
\partial_{_{z_{_1}^a}}\partial_{_{z_{_2}^a}}\partial_{_{z_{_3}^a}}-\partial_{_{z_{_4}^a}}\partial_{_{z_{_5}^a}}\partial_{_{z_{_6}^a}},\;
\partial_{_{z_{_1}^a}}\partial_{_{z_{_7}^a}}\partial_{_{z_{_8}^a}}-\partial_{_{z_{_2}^a}}\partial_{_{z_{_3}^a}}\partial_{_{z_{_6}^a}}\}\;.
\label{GKZ-1M-10}
\end{eqnarray}
Certainly one can calculate the state polytope~\cite{Sturmfels1995} corresponding to the universal Gr\"obner basis
${\cal U}_{_{\mathbf{A}_a}}$ whose normal fan coincides with the Gr\"obner fan, then construct
canonical series solutions in the convergent regions which are presented in Ref.~\cite{Feng2019}.
In order to perform the analytic continuation of canonical series solutions from the convergent regions
to the whole parameter space, one utilizes some linear fractional transformations
among the complex variables $z_{_{1}}^a,\cdots,z_{_{8}}^a$, then chooses $u_{_k}=1,\;v_{_j}=1$
$(k=1,2,3,\;j=1,2)$ finally.

\subsection{$C_{_0}$ function with three differential masses\label{sec4-3}}
\indent\indent
The massive $C_{_0}$ function is generally written as
\begin{eqnarray}
&&C_{_0}(p_{_1}^2,\;p_{_2}^2,\;p_{_3}^2,\;m_{_1}^2,\;m_{_2}^2,\;m_{_3}^2)
\nonumber\\
&&\hspace{-0.5cm}=
\int{d^Dq\over(2\pi)^D}{1\over(q^2-m_{_3}^2)((q+p_{_1})^2-m_{_2}^2)
((q-p_{_2})^2-m_{_1}^2)}
\nonumber\\
&&\hspace{-0.5cm}=
{i(p_{_3}^2)^{D/2-3}\over(4\pi)^{D/2}}F_{_{p_{_3}}}({\bf a},\;{\bf b};\;\;{\bf x})
\label{GKZ-3M-1}
\end{eqnarray}
with the row vectors ${\bf a}=(a_{_1},a_{_2},a_{_3})\in {\bf R}^3$, ${\bf b}=(b_{_1},b_{_2})\in{\bf R}^2$,
and ${\bf x}=(\xi_{_{13}},\xi_{_{23}},\xi_{_{33}},x_{_{13}},x_{_{23}})\in{\bf R}^5$, respectively.
Here the parameters $a_{_k},\;b_{_j}\;(k=1,2,3,\;j=1,2)$ are taken the same values as in Eq.~(\ref{GKZ-1M-1})
and the dimensionless ratios $\xi_{_{ij}}=-m_{_i}^2/p_{_j}^2$, $x_{_{ij}}=p_{_i}^2/p_{_j}^2$.
Furthermore the dimensionless function $F_{_{p_{_3}}}$ complies with the hypergeometric system of linear PDEs
\begin{eqnarray}
&&\Big\{\Big[3-{D\over2}+\sum\limits_{i=1}^3\hat{\vartheta}_{_{\xi_{_{i3}}}}
+\sum\limits_{i=1}^2\hat{\vartheta}_{_{x_{_{i3}}}}\Big]\Big[1+\hat{\vartheta}_{_{\xi_{_{33}}}}
+\sum\limits_{i=1}^2\hat{\vartheta}_{_{x_{_{i3}}}}\Big]
\Big[4-D+\sum\limits_{i=1}^3\hat{\vartheta}_{_{\xi_{_{i3}}}}\Big]
\nonumber\\
&&\hspace{0.0cm}
-{1\over\xi_{_{33}}}\hat{\vartheta}_{_{\xi_{_{33}}}}
\prod\limits_{i=1}^3\Big[2-{D\over2}+\hat{\vartheta}_{_{\xi_{_{i3}}}}+\hat{\vartheta}_{_{\xi_{_{33}}}}
+\hat{\vartheta}_{_{x_{_{(3-i)3}}}}\Big]\Big\}F_{_{p_{_3}}}=0
\;,\nonumber\\
&&\Big\{\Big[3-{D\over2}+\sum\limits_{i=1}^3\hat{\vartheta}_{_{\xi_{_{i3}}}}
+\sum\limits_{i=1}^2\hat{\vartheta}_{_{x_{_{i3}}}}\Big]
\Big[4-D+\sum\limits_{i=1}^3\hat{\vartheta}_{_{\xi_{_{i3}}}}\Big]
\nonumber\\
&&\hspace{0.0cm}
+{1\over\xi_{_{j3}}}\hat{\vartheta}_{_{\xi_{_{j3}}}}
\Big[2-{D\over2}+\hat{\vartheta}_{_{\xi_{_{33}}}}+\hat{\vartheta}_{_{\xi_{_{j3}}}}
+\hat{\vartheta}_{_{x_{_{(3-j)3}}}}\Big]\Big\}F_{_{p_{_3}}}=0
\;,\nonumber\\
&&\Big\{\Big[3-{D\over2}+\sum\limits_{i=1}^3\hat{\vartheta}_{_{\xi_{_{i3}}}}
+\sum\limits_{i=1}^2\hat{\vartheta}_{_{x_{_{i3}}}}\Big]\Big[1+\hat{\vartheta}_{_{\xi_{_{33}}}}
+\sum\limits_{i=1}^2\hat{\vartheta}_{_{x_{_{i3}}}}\Big]
\nonumber\\
&&\hspace{0.0cm}
-{1\over x_{_{j3}}}\hat{\vartheta}_{_{x_{_{j3}}}}
\Big[2-{D\over2}+\hat{\vartheta}_{_{\xi_{_{33}}}}+\hat{\vartheta}_{_{\xi_{_{(3-j)3}}}}
+\hat{\vartheta}_{_{x_{_{j3}}}}\Big]\Big\}F_{_{p_{_3}}}=0,\;(j=1,\;2)\;.
\label{GKZ-3M-2}
\end{eqnarray}
Defining Miller's transformation on the function $F_{_{p_{_3}}}$
\begin{eqnarray}
&&\Phi_{_{p_{_3}}}({\bf a},\;{\bf b};\;\;{\bf x},\;{\bf u},\;{\bf v})
={\bf u}^{\bf a}{\bf v}^{{\bf b}-{\bf e}_{_2}}
F_{_{p_{_3}}}({\bf a},\;{\bf b};\;\;{\bf x})
\label{GKZ-3M-3}
\end{eqnarray}
with ${\bf u}=(u_{_1},u_{_2},u_{_3})\in{\bf R}^3$, ${\bf v}=(v_{_1},v_{_2})\in{\bf R}^2$,
we similarly replace the multiplication of the parameters $a_{_k},\;b_{_j}$ by Euler operators
$\hat{\vartheta}_{_{u_{_k}}}$, $\hat{\vartheta}_{_{v_{_j}}}$:
\begin{eqnarray}
&&\hat{\vartheta}_{_{u_{_k}}}\Phi_{_{p_{_3}}}({\bf a},\;{\bf b};\;\;{\bf x},\;{\bf u},\;{\bf v})=
a_{_k}\Phi_{_{p_{_3}}}({\bf a},\;{\bf b};\;\;{\bf x},\;{\bf u},\;{\bf v})
\;,\nonumber\\
&&\hat{\vartheta}_{_{v_{_j}}}\Phi_{_{p_{_3}}}({\bf a},\;{\bf b};\;\;{\bf x},\;{\bf u},\;{\bf v})=
(b_{_j}-1)\Phi_{_{p_{_3}}}({\bf a},\;{\bf b};\;\;{\bf x},\;{\bf u},\;{\bf v})
\label{GKZ-3M-4}
\end{eqnarray}
where $k=1,\;2,\;3$ and $j=1,\;2$ respectively.
Similarly the contiguous relations of the function defined in Eq.~(\ref{GKZ-3M-3}) are
\begin{eqnarray}
&&\hat{\cal O}_{_k}\Phi_{_{p_{_3}}}({\bf a},\;{\bf b};\;\;{\bf x},\;{\bf u},\;{\bf v})=
a_{_{k}}\Phi_{_{p_{_3}}}({\bf a}+{\bf n}_{_{3,k}},\;{\bf b};\;\;{\bf x},\;{\bf u},\;{\bf v})
\;,\nonumber\\
&&\hat{\cal O}_{_{3+j}}\Phi_{_{p_{_3}}}({\bf a},\;{\bf b};\;\;{\bf x},\;{\bf u},\;{\bf v})=
(b_{_j}-1)\Phi_{_{p_{_3}}}({\bf a},\;{\bf b}-{\bf n}_{_{2,j}};\;\;{\bf x},\;{\bf u},\;{\bf v})
\;,\nonumber\\
&&\hat{\cal O}_{_{5+j}}\Phi_{_{p_{_3}}}({\bf a},\;{\bf b};\;\;{\bf x},\;{\bf u},\;{\bf v})=
{a_{_{2}}a_{_{3}}\over b_{_{j}}}\Phi_{_{p_{_3}}}({\bf a}+{\bf n}_{_{3,1}}
+{\bf n}_{_{3,2}},\;{\bf b}+{\bf n}_{_{2,j}};\;\;{\bf x},\;{\bf u},\;{\bf v})
\;,\nonumber\\
&&\hat{\cal O}_{_8}\Phi_{_{p_{_3}}}({\bf a},\;{\bf b};\;\;{\bf x},\;{\bf u},\;{\bf v})=
{a_{_{1}}a_{_{2}}a_{_{3}}\over b_{_{1}}b_{_{2}}}
\Phi_{_{p_{_3}}}({\bf a}+{\bf n}_{_3},\;{\bf b}+{\bf n}_{_2};\;\;{\bf x},\;{\bf u},\;{\bf v})
\;,\nonumber\\
&&\hat{\cal O}_{_{8+j}}\Phi_{_{p_{_3}}}({\bf a},\;{\bf b};\;\;{\bf x},\;{\bf u},\;{\bf v})=
{a_{_{2}}a_{_{3}}\over b_{_{j}}}\Phi_{_{p_{_3}}}({\bf a}+{\bf n}_{_{3,2}}
+{\bf n}_{_{3,3}},\;{\bf b}+{\bf n}_{_{2,j}};\;\;{\bf x},\;{\bf u},\;{\bf v})\;,
\label{GKZ-3M-5}
\end{eqnarray}
where the operators $\hat{\cal O}_{_n}(n=1,\cdots,10)$ are defined as
\begin{eqnarray}
&&\hat{\cal O}_{_1}=u_{_1}(\sum\limits_{i=1}^3\hat{\vartheta}_{_{\xi_{_{i3}}}}+\hat{\vartheta}_{_{u_{_1}}})
\;,\nonumber\\
&&\hat{\cal O}_{_2}=u_{_2}(\sum\limits_{i=1}^3\hat{\vartheta}_{_{\xi_{_{i3}}}}
+\sum\limits_{i=1}^2\hat{\vartheta}_{_{x_{_{i3}}}}+\hat{\vartheta}_{_{u_{_2}}})
\;,\nonumber\\
&&\hat{\cal O}_{_3}=u_{_3}(\hat{\vartheta}_{_{\xi_{_{33}}}}
+\sum\limits_{i=1}^2\hat{\vartheta}_{_{x_{_{i3}}}}+\hat{\vartheta}_{_{u_{_3}}})
\;,\nonumber\\
&&\hat{\cal O}_{_{3+j}}={1\over v_{_j}}(\sum\limits_{i=1}^3\hat{\vartheta}_{_{\xi_{_{i3}}}}-\hat{\vartheta}_{_{\xi_{_{j3}}}}
+\hat{\vartheta}_{_{x_{_{j3}}}}+\hat{\vartheta}_{_{v_{_j}}})\;\;(j=1,\;2)
\;,\nonumber\\
&&\hat{\cal O}_{_{6}}=u_{_1}u_{_2}v_{_1}\partial_{_{\xi_{_{23}}}}
\;,\nonumber\\
&&\hat{\cal O}_{_{7}}=u_{_1}u_{_2}v_{_2}\partial_{_{\xi_{_{13}}}}
\;,\nonumber\\
&&\hat{\cal O}_{_{8}}=u_{_1}u_{_2}u_{_3}v_{_1}v_{_2}\partial_{_{\xi_{_{33}}}}
\;,\nonumber\\
&&\hat{\cal O}_{_{8+j}}=u_{_2}u_{_3}v_{_j}\partial_{_{x_{_{j3}}}}
\;,(j=1,\;2)\;.
\label{GKZ-3M-6}
\end{eqnarray}
Those operators together with $\hat{\vartheta}_{_{u_{_i}}},\;\hat{\vartheta}_{_{v_{_j}}}$ define the Lie algebra of the
hypergeometric system~\cite{Miller68,Miller72} in Eq.(\ref{GKZ-3M-2}). Through the transformation of indeterminates
\begin{eqnarray}
&&z_{_i}={1\over u_{_i}}\;,\;\;z_{_{3+j}}=v_{_j}\;,\;\;
z_{_{6}}={\xi_{_{13}}\over u_{_1}u_{_2}v_{_2}}\;,\;\;z_{_{7}}={\xi_{_{23}}\over u_{_1}u_{_2}v_{_1}}\;,
\nonumber\\
&&z_{_{8}}={\xi_{_{33}}\over u_{_1}u_{_2}u_{_3}v_{_1}v_{_2}}\;,
z_{_{8+j}}={x_{_{j3}}\over u_{_2}u_{_3}v_{_j}}\;,\;\;(i=1,\;2,\;3,\;\;j=1,\;2)\;,
\label{GKZ-3M-7}
\end{eqnarray}
the equations in Eq.~(\ref{GKZ-3M-4}) are
\begin{eqnarray}
&&\Big(\mathbf{A}\cdot\vec{\vartheta}\Big)\Phi_{_{p_{_3}}}({\bf a},\;{\bf b};\;\;{\bf x},\;{\bf u},\;{\bf v})
=\mathbf{B}\Phi_{_{p_{_3}}}({\bf a},\;{\bf b};\;\;{\bf x},\;{\bf u},\;{\bf v})\;,
\label{GKZ-3M-8}
\end{eqnarray}
where
\begin{eqnarray}
&&\mathbf{A}=\left(\begin{array}{cccccccccc}1\;&0\;&0\;&0\;&0\;&1\;&1\;&1\;&0\;&0\;\\0\;&1\;&0\;&0\;&0\;&1\;&1\;&1\;&1\;&1\;\\
0\;&0\;&1\;&0\;&0\;&0\;&0\;&1\;&1\;&1\;\\0\;&0\;&0\;&1\;&0\;&0\;&-1\;&-1\;&-1\;&0\;\\0\;&0\;&0\;&0\;&1\;&-1\;&0\;&-1\;&0\;&-1\;
\end{array}\right)
\;,\nonumber\\
&&\vec{\vartheta}^{\;T}=(\vartheta_{_{z_{_1}}},\;\vartheta_{_{z_{_2}}},\;\vartheta_{_{z_{_3}}},\;
\vartheta_{_{z_{_4}}},\;\vartheta_{_{z_{_5}}},\;\vartheta_{_{z_{_6}}},\;\vartheta_{_{z_{_7}}},\;
\vartheta_{_{z_{_8}}},\;\vartheta_{_{z_{_9}}},\;\vartheta_{_{z_{_{10}}}})
\;,\nonumber\\
&&\mathbf{B}^{\;T}=(-a_{_1},\;-a_{_2},\;-a_{_3},\;b_{_1},\;b_{_2})\;.
\label{GKZ-3M-9}
\end{eqnarray}
Correspondingly the universal Gr\"obner basis of the toric ideal associated with $\mathbf{A}$ is
\begin{eqnarray}
&&{\cal U}_{_\mathbf{A}}=\{\partial_{_{z_{_1}}}\partial_{_{z_{_2}}}-\partial_{_{z_{_4}}}\partial_{_{z_{_7}}},\;
\partial_{_{z_{_1}}}\partial_{_{z_{_2}}}-\partial_{_{z_{_5}}}\partial_{_{z_{_6}}},\;
\partial_{_{z_{_1}}}\partial_{_{z_{_9}}}-\partial_{_{z_{_3}}}\partial_{_{z_{_7}}},\;
\partial_{_{z_{_1}}}\partial_{_{z_{_9}}}-\partial_{_{z_{_5}}}\partial_{_{z_{_8}}},\;
\nonumber\\
&&\hspace{1.2cm}
\partial_{_{z_{_1}}}\partial_{_{z_{_{10}}}}-\partial_{_{z_{_3}}}\partial_{_{z_{_6}}},\;
\partial_{_{z_{_1}}}\partial_{_{z_{_{10}}}}-\partial_{_{z_{_4}}}\partial_{_{z_{_8}}},\;
\partial_{_{z_{_2}}}\partial_{_{z_{_3}}}-\partial_{_{z_{_4}}}\partial_{_{z_{_9}}},\;
\partial_{_{z_{_2}}}\partial_{_{z_{_3}}}-\partial_{_{z_{_5}}}\partial_{_{z_{_{10}}}},\;
\nonumber\\
&&\hspace{1.2cm}
\partial_{_{z_{_2}}}\partial_{_{z_{_8}}}-\partial_{_{z_{_6}}}\partial_{_{z_{_9}}},\;
\partial_{_{z_{_2}}}\partial_{_{z_{_8}}}-\partial_{_{z_{_7}}}\partial_{_{z_{_{10}}}},\;
\partial_{_{z_{_3}}}\partial_{_{z_{_6}}}-\partial_{_{z_{_4}}}\partial_{_{z_{_{8}}}},\;
\partial_{_{z_{_3}}}\partial_{_{z_{_7}}}-\partial_{_{z_{_5}}}\partial_{_{z_{_{8}}}},\;
\nonumber\\
&&\hspace{1.2cm}
\partial_{_{z_{_4}}}\partial_{_{z_{_7}}}-\partial_{_{z_{_5}}}\partial_{_{z_{_{6}}}},\;
\partial_{_{z_{_4}}}\partial_{_{z_{_9}}}-\partial_{_{z_{_5}}}\partial_{_{z_{_{10}}}},\;
\partial_{_{z_{_6}}}\partial_{_{z_{_9}}}-\partial_{_{z_{_7}}}\partial_{_{z_{_{10}}}},\;
\partial_{_{z_{_1}}}\partial_{_{z_{_2}}}\partial_{_{z_{_3}}}-\partial_{_{z_{_4}}}\partial_{_{z_{_5}}}\partial_{_{z_{_{8}}}},\;
\nonumber\\
&&\hspace{1.2cm}
\partial_{_{z_{_1}}}\partial_{_{z_{_2}}}\partial_{_{z_{_8}}}-\partial_{_{z_{_3}}}\partial_{_{z_{_6}}}\partial_{_{z_{_{7}}}},\;
\partial_{_{z_{_1}}}\partial_{_{z_{_2}}}\partial_{_{z_{_9}}}-\partial_{_{z_{_5}}}\partial_{_{z_{_7}}}\partial_{_{z_{_{10}}}},\;
\partial_{_{z_{_1}}}\partial_{_{z_{_2}}}\partial_{_{z_{_{10}}}}-\partial_{_{z_{_4}}}\partial_{_{z_{_6}}}\partial_{_{z_{_{9}}}},\;
\nonumber\\
&&\hspace{1.2cm}
\partial_{_{z_{_1}}}\partial_{_{z_{_4}}}\partial_{_{z_{_{9}}}}-\partial_{_{z_{_3}}}\partial_{_{z_{_5}}}\partial_{_{z_{_{6}}}},\;
\partial_{_{z_{_1}}}\partial_{_{z_{_5}}}\partial_{_{z_{_{10}}}}-\partial_{_{z_{_3}}}\partial_{_{z_{_4}}}\partial_{_{z_{_{7}}}},\;
\partial_{_{z_{_1}}}\partial_{_{z_{_6}}}\partial_{_{z_{_{9}}}}-\partial_{_{z_{_4}}}\partial_{_{z_{_7}}}\partial_{_{z_{_{8}}}},\;
\nonumber\\
&&\hspace{1.2cm}
\partial_{_{z_{_1}}}\partial_{_{z_{_7}}}\partial_{_{z_{_{10}}}}-\partial_{_{z_{_5}}}\partial_{_{z_{_6}}}\partial_{_{z_{_{8}}}},\;
\partial_{_{z_{_1}}}\partial_{_{z_{_9}}}\partial_{_{z_{_{10}}}}-\partial_{_{z_{_2}}}\partial_{_{z_{_3}}}\partial_{_{z_{_{8}}}},\;
\partial_{_{z_{_2}}}\partial_{_{z_{_3}}}\partial_{_{z_{_{6}}}}-\partial_{_{z_{_4}}}\partial_{_{z_{_7}}}\partial_{_{z_{_{10}}}},\;
\nonumber\\
&&\hspace{1.2cm}
\partial_{_{z_{_2}}}\partial_{_{z_{_3}}}\partial_{_{z_{_{7}}}}-\partial_{_{z_{_5}}}\partial_{_{z_{_6}}}\partial_{_{z_{_{9}}}},\;
\partial_{_{z_{_2}}}\partial_{_{z_{_4}}}\partial_{_{z_{_{8}}}}-\partial_{_{z_{_5}}}\partial_{_{z_{_6}}}\partial_{_{z_{_{10}}}},\;
\partial_{_{z_{_2}}}\partial_{_{z_{_5}}}\partial_{_{z_{_{8}}}}-\partial_{_{z_{_4}}}\partial_{_{z_{_7}}}\partial_{_{z_{_{9}}}},\;
\nonumber\\
&&\hspace{1.2cm}
\partial_{_{z_{_3}}}\partial_{_{z_{_6}}}\partial_{_{z_{_{9}}}}-\partial_{_{z_{_5}}}\partial_{_{z_{_8}}}\partial_{_{z_{_{10}}}},\;
\partial_{_{z_{_3}}}\partial_{_{z_{_7}}}\partial_{_{z_{_{10}}}}-\partial_{_{z_{_4}}}\partial_{_{z_{_8}}}\partial_{_{z_{_{9}}}}\}\;.
\label{GKZ-3M-10}
\end{eqnarray}
The normal fan of corresponding state polytope of the universal Gr\"obner basis
is the Gr\"obner fan, canonical series solutions are obtained similarly
in the convergent regions.
To perform the analytic continuation of canonical series solutions from the convergent regions
to the whole parameter space, one can use the linear fractional transformations
among the complex variables $z_{_{1}},\cdots,z_{_{10}}$, then sets $u_{_k}=1,\;v_{_j}=1$
$(k=1,2,3,\;j=1,2)$ finally.

\section{Summary\label{sec5}}
\indent\indent
Using the system of linear PDEs satisfied by the corresponding Feynman integral
in Refs.~\cite{Feng2018,Feng2019}, we present GKZ-hypergeometric systems
of one-loop self energy, one-loop triangle, two-loop vacuum,
and the two-loop sunset diagrams, respectively.
In those GKZ-hypergeometric systems the codimension equals the number of independent
dimensionless ratios among the external momentum squared and virtual mass squared.

Actually one can derive GKZ-hypergeometric systems from Mellin-Barnes representations
for the one-loop Feynman diagrams and those multiloop diagrams with two vertices,
whose codimension equals the number of independent dimensionless ratios among the external momentum squared
and virtual mass squared. Nevertheless for the generic multiloop Feynman diagrams,
the corresponding codimension of GKZ-hypergeometric system is far larger than the number of independent
dimensionless ratios, whether using Mellin-Barnes or Lee-Pomeransky representations. In order to construct
canonical series solutions properly, the corresponding GKZ-hypergeometric
system should be restricted to the hyperplane in parameter space.

Taking GKZ-hypergeometric systems of one-loop self energy, one-loop massless triangle,
and two-loop vacuum diagrams as examples, we present in detail how to perform triangulation
and how to construct canonical series solutions in the corresponding convergent regions. The analytic
continuation of those series solutions is performed through some well known reduction of
Appell function of the fourth kind. In order to make analytic continuation of those
series solutions of GKZ-hypergeometric systems of the massive sunset and
massive one-loop triangle diagrams etc., one can perform the linear fractional transformations
among the complex variables.

One of the techniques not involved here is how to project GKZ-hypergeometric system
to the restricting hyperplane. Another calculation not contained here is how to make analytic continuation of those
canonical series solutions from their convergent regions to the whole parameter space through
linear fractional transformation among the complex variables.
Algorithm for the first problem has been presented in literature already,
and the second problem is attributed to a problem of integer programming~\cite{Schrijver1999} in principle.
We will release our results relating to those topics in near future elsewhere.

\begin{acknowledgments}
\indent\indent
The work has been supported partly by the National Natural
Science Foundation of China (NNSFC) with Grant No. 11535002, No. 11821505,
No. 11447601, No. 11675239, No. 11705045, and No. 11805140,
and the youth top-notch talent support program of Hebei province.
Furthermore, the author (C.-H. Chang) is also supported by Key Research
program of Frontier Sciences, CAS, Grant No. QYZDY-SSW-SYS006.
\end{acknowledgments}



\begin{thebibliography}{99}
\bibitem{CEPC-SPPC}CEPC-SPPC study group,
{\it CEPC-SPPC preliminary conceptual design report. 1. Physics and detector}, IHEP-CEPC-DR-2015-01, 2015.
\bibitem{ILC}T.~Behnke et al., {\it The International Linear Collider Technical Design Report - Volume I: Executive Summary},
arXiv:1306.6327 [physics.acc-ph].
\bibitem{HI-LHC}G.~Apollinari, et al., {\it High-Luminosity Large Hadron Collider (HL-LHC):
Preliminary Design Report}, Technical Report CERN-2015-005 (2015).
\bibitem{CMS2012}{\rm CMS}~Collaboration, Phys.~Lett.~B{\bf 716}(2012)30.
\bibitem{ATLAS2012}{\rm  ATLAS}~Collaboration, Phys.~Lett.~B{\bf716}(2012)1.

\bibitem{tHooft1979}G.~t'Hooft, M.~Veltman, Nucl.~Phys.~B{\bf153}(1979)365.
\bibitem{Passarino1979}G.~Passarino, M.~Veltman, Nucl.~Phys.~B{\bf160}(1979)151.
\bibitem{V.A.Smirnov2012}V.~A.~Smirnov, {\it Analytic Tools for Feynman Integrals},
(Springer, Heidelberg 2012), and references therein.

\bibitem{Regge1967}T.~Regge, {\it Algebraic Topology Methods in the Theory of Feynman Relativistic
Amplitudes}, In: {\it Battelle Rencontres - 1967 Lectures in Math. and Phys.,} edited by C.~M.~DeWitt
and J.~A.~Wheeler, 1967, pp.~433-458.
\bibitem{Kashiwara1976}M.~Kashiwara, T.~Kawai, {\it Holonomic Systems of Linear Differential Equations
and Feynman Integrals}, Publications of the Research Institute for Math. Sci. {\bf 12}(1976)131.
\bibitem{Nasrollahpoursamami2016}E.~Nasrollahpoursamami, {\it Periods of Feynman Diagrams and GKZ D-modules},
arXiv:1605.04970~[math-ph].

\bibitem{Gelfand1987}I.~M.~Gel'fand, Soviet~Math.~Dokl.~{\bf 33}(1986)573.
\bibitem{Gelfand1988}I.~M.~Gel'fand, M.~I.~Graev, and A.~V.~Zelevinsky, Soviet~Math.~Dokl.~{\bf 36}(1988)5.
\bibitem{Gelfand1988a}I.~M.~Gel'fand, A.~V.~Zelevinsky, and M.~M.~Kapranov, Soviet~Math.~Dokl.~{\bf 37}(1988)678.
\bibitem{Gelfand1989}I.~M.~Gel'fand, A.~V.~Zelevinsky, and M.~M.~Kapranov, Funct.~Anal.~Appl.~{\bf 23}(1989)94.
\bibitem{Gelfand1990}I.~M.~Gel'fand, M.~M.~Kapranov, and A.~V.~Zelevinsky, Adv.~in~Math.~{\bf 84}(1990)255.

\bibitem{Davydychev1}E.~E.~Boos, and A.~I.~Davydychev, Vestn.~Mosk.~Univ.~{\bf28}(1987)8.
\bibitem{Davydychev1993NPB}A.~I.~Davydychev, and J.~B.~Tausk, Nucl.~Phys.~B{\bf397}(1993)123.
\bibitem{Davydychev2000}A.~I.~Davydychev, Phys.~Rev.~D{\bf61}(2000)087701.
\bibitem{Davydychev3}E.~E.~Boos, and A.~I.~Davydychev, Theor.~Math.~Phys.~{\bf89}(1991)1052.
\bibitem{Davydychev1992JPA}A.~I.~Davydychev, J.~Phys.~A{\bf25}(1992)5587.
\bibitem{Davydychev1992JMP}A.~I.~Davydychev, J.~Math.~Phys.{\bf33}(1992)358.
\bibitem{Davydychev1991JMP}A.~I.~Davydychev, J.~Math.~Phys.{\bf32}(1991)1052.
\bibitem{Davydychev1993}N.~I.~Ussyukina, and A.~I.~Davydychev, Phys.~Lett.~B{\bf298}(1993)363.
\bibitem{Davydychev2006}A.~I.~Davydychev, Nucl.~Instrum.~Meth.~A{\bf559}(2006)293.
\bibitem{Tarasov2000}O.~V.~Tarasov, Nucl.~Phys.~B(Proc.~Suppl){\bf 89}(2000)237.
\bibitem{Tarasov2003}J.~Fleischer, F.~Jegerlehner, and O.~V.~Tarasov, Nucl.~Phys.~B{\bf 672}(2003)303.

\bibitem{Kalmykov2009}M.~Yu.~Kalmykov, B.~A.~Kniehl, Nucl.~Phys.~B{\bf 809}(2009)365.
\bibitem{Bytev2010}V.~V.~Bytev, M.~Yu.~Kalmykov, B.~A.~Kniehl, Nucl.~Phys.~B{\bf 836}(2010)129.
\bibitem{Kalmykov2011}M.~Yu.~Kalmykov, B.~A.~Kniehl, Phys.~Lett.~B{\bf 702}(2011)268.
\bibitem{Bytev2013}V.~V.~Bytev, M.~Yu.~Kalmykov, B.~A.~Kniehl, Comput.~Phys.~Commun{\bf 184}(2013)2332.
\bibitem{Bytev2015}V.~V.~Bytev, M.~Yu.~Kalmykov, Comput.~Phys.~Commun{\bf 189}(2015)128.
\bibitem{Bytev2016}V.~V.~Bytev, M.~Yu.~Kalmykov, Comput.~Phys.~Commun{\bf 206}(2016)78.
\bibitem{Kalmykov2012}M.~Y.~Kalmykov, and B.~A.~Kniehl, Phys.~Lett.~B{\bf 714}(2012)103.
\bibitem{Kalmykov2017}M.~Y.~Kalmykov, and B.~A.~Kniehl, JHEP{\bf 1707}(2017)031.


\bibitem{Cruz2019}L.~Cruz, {\it Feynman integrals as A-hypergeometric functions},
arXiv:1907.00507~[math-ph].
\bibitem{Lee2013}R.~N.~Lee, A.~A.~Pomeransky, JHEP{\bf 1311}(2013)165.
\bibitem{Klausen2019}R.~Klausen, {\it Hypergeometric Series Representations of Feynman
integrals by GKZ Hypergeometric Systems}, arXiv:1910.08651~[hep-th].
\bibitem{Oaku1997}T.~Oaku, Adv.~in~Appl.~Math.~{\bf19}(1997)61.
\bibitem{Walther1999}U.~Walther, J.~of~Pure~and~Applied~Algebra.~{\bf139}(1999)303.
\bibitem{Oaku2001}T.~Oaku, N.~Takayama, J.~of~Pure~and~Applied~Algebra.~{\bf156}(2001)267.

\bibitem{Feng2018}T.-F.~Feng, C.-H.~Chang, J.-B.~Chen, Z.-H.~Gu, and H.-B.~Zhang,
Nucl.~Phys.~B{\bf927}(2018)516.
\bibitem{Feng2019}T.-F.~Feng, C.-H.~Chang, J.-B.~Chen, and H.-B.~Zhang,
Nucl.~Phys.~B{\bf940}(2019)130.

\bibitem{Miller68}W.~Miller~Jr., J.~Math.~Mech.~{\bf17}(1968)1143.
\bibitem{Miller72}W.~Miller~Jr., SIAM.~J.~Math.~Anal.~{\bf3}(1972)31.
\bibitem{M.Saito2000}M.~Saito, B.~Sturmfels, N.~Takayama, {\it Gr\"obner Deformations
of Hypergeometric Differential Equations}, Springer 2000.
\bibitem{Horn1889}J.~Horn, Math.~Ann.~{\bf34}(1889)544.

\bibitem{M.E.Taylor12}M.~E.~Taylor, {\it Partial differential equations}
(Springer, Heidelberg 2012).
\bibitem{L.J.Slater66}L.~J.~Slater, {\it Generalised Hypergeometric Functions}
(Cambridge University Press 1966).
\bibitem{J.Leray1959}J.~Leray, {\it Le Calcul Diff\'erential et Int\'egral sur Une Vari\'et\'e
Analytique Complexe}, Bull.~Soc.~Math., {\bf 87}(1959)81.

\bibitem{Cox1991}D.~Cox, J.~Little, D.~O'Shea, {\it Ideals, Varieties and Algorithms}
(Springer, New York 1991).
\bibitem{Cox1998}D.~Cox, J.~Little, D.~O'Shea, {\it Using Algebraic Geometry}
(Springer, New York 1998).
\bibitem{Sturmfels1995}B.~Sturmfels, {\it Gr\"obner Bases and Convex Polytopes},
University Lecture Notes, Vol.~8.~American Mathematical Society, Providence 1995.
\bibitem{Eisenbud1995}D.~Eisenbud, {\it Commutative Algebra with a View Toward Algebraic Geometry}
(Springer, New York 1995).
\bibitem{Coutinho1995}S.~C.~Coutinho, {\it A Primer of Algebraic D-modules},
London Mathematical Society Student Text 33, Cambridge Univ. Press, Cambridge 1995.

\bibitem{Vanhove2018}P.~Vanhove, {\it Feynman integrals, toric geometry and mirror symmetry},
arXiv:1807.11466~[hep-th].

\bibitem{V.A.Smirnov1999}V.~A.~Smirnov, Phys.~Lett.~B{\bf460}(1999)397.
\bibitem{J.B.Tausk1999}J.~B.~Tausk,  Phys.~Lett.~B{\bf469}(1999)225.
\bibitem{Gelfand1994}I.~M.~Gel'fand, M.~M.~Kapranov, and A.~V.~Zelevinsky,
{\it Discriminants, Resultants and Multidimensional Determinants}, Birkh\"auser, Boston 1994.

\bibitem{Berends1994}F.~A.~Berends, M.~B\"ohm, M.~Buza, and R.~Scharf, Z.~Phys.~C{\bf63}(1994)227.
\bibitem{Ananthanarayan2019}B.~Ananthanarayan, S.~Friot, and S.~Ghosh, {\it New series representations
for the two-loop massive sunset diagram}, arXiv:1911.10096~[hep-ph].

\bibitem{Schrijver1999}A.~Schrijver, {\it Theory of Linear and Integer Programming},
(John Wiley 1998).


\end{thebibliography}
\end{document}